\documentclass[twocolumn]{aastex631}
\usepackage{microtype}  
\usepackage{placeins}    
\usepackage{enumitem}

\usepackage[hyphens]{url}
\usepackage{amsmath}
\usepackage{hyperref}
\begin{document}

\title{An optical--mid-infrared color evolution tool for nova identification using WISE data}

\author[0000-0001-8784-6436]{Joseph Onuegbu}
\affil{Department of Physics, Ariel University, Ariel, Israel}
\author[0000-0002-7349-1109]{Dafne Guetta}
\affil{Department of Physics, Ariel University, Ariel, Israel}
\author[0000-0002-0023-0485]{Yael Hillman}
\affil{Department of Physics, Azrieli College of Engineering, Jerusalem, Israel}
\author{Volker Perdelwitz}
\affil{Department of Earth and Planetary Science, Weizmann Institute of Science, Rehovot, Israel}
\author{Massimo Della Valle}
\affil{INAF-Capodimonte, Napoli}
\affil{Department of Physics, Ariel University, Ariel, Israel}

\begin{abstract}
We present a novel approach for characterizing nova candidates by exploiting the infrared capabilities of the Wide-field Infrared Survey Explorer (WISE) catalog. We developed a pipeline to identify novae based on well-defined infrared criteria, and leveraging this pipeline, we successfully identified 41 optically confirmed novae in the WISE catalog. In particular, we focus on the color difference between the optical $V$ band and the WISE 3.4$\mu$m W1 band as a diagnostic. We compared their infrared light curves with their optical counterparts. We identified a strong correlation from which we proposed a color difference model ($V-W1$) that can be used for further identification and characterization of novae.
Our analysis validates the mass-loss timescale theory, which predicts that systems with lower accretion rates accumulate larger envelopes and produce more massive ejecta. We also confirm models' prediction that the early color evolution of novae is governed by ejecta expansion and cooling. From our sample statistics, we infer a Galactic nova rate of $\sim$ 40–50 novae/yr, consistent with modern and infrared-corrected estimates.
The resultant model from this work paves the way for future large-scale investigations of nova candidates.  

\end{abstract}


\section{Introduction} \label{sec:intro}
\subsection{Novae}
Nova eruptions occur in compact binaries where a white dwarf (WD) accretes hydrogen-rich material from a companion star \cite[]{1978ARA&A..16..171G,2020A&ARv..28....3D}.
Mass transfer from the companion star leads to the accumulation of material on the WD's surface \citep{1998PASP..110....3G, Starrfield_2016, 2021MNRAS.505.3260H}. 
WD envelopes become increasingly electron-degenerate as their masses increase. According to \cite{1972ApJ...176..169S}, as the accreted material accumulates on the surface of the WD, it is compressed and heated, gradually forming a degenerate layer. Over time, this layer increases in temperature and density, leading to hydrogen fusion reactions occurring at an accelerating rate. 
A thermonuclear runaway (TNR) occurs when the energy generated by hydrogen fusion on the white dwarf’s surface can no longer be dissipated efficiently. This leads to an exponential increase in temperature and fusion rate, causing the nova to reach a maximum absolute magnitude typically between M $\approx$ -5 and M $\approx$ -10 \citep{1978A&A....62..339P, 1986ApJ...310..222P, 2009ApJ...690.1148S, Starrfield_2016, 2016ApJ...833..149D}.  
 The WD’s electron degeneracy is lifted due to the intense heat and pressure generated by the fusion reactions, and the overlying envelope expands rapidly \citep{1982ApJ...261..649S, 1991ApJ...376..177R, 1994ApJ...424..319K, 2022MNRAS.511.5570H}. 
 
 The degenerate equation of state of a WD ensures that its radius decreases and its gravitational potential increases greatly as its mass increases.\citep{1931ApJ....74...81C, 1935MNRAS..95..207C, 2016ApJ...819..168H, Maguire_2020}.  Consequently, the accreted layer expands and is then ejected, either partially or completely, possibly with additional material from the WD's outer layers \citep{1995ApJ...445..789P,1998PASP..110....3G,2005ApJ...623..398Y,2016ApJ...819..168H,2020NatAs...4..886H,2020ApJ...895...70S,2022MNRAS.515.1404H}. 
 This process results in a dramatic brightening of the system, lasting from weeks to years \citep{1989clno.conf.....B, 2021ARA&A..59..391C}.

The accretion rate and the WD mass are key factors that determine the characteristics of the nova eruption, a realization that emerged from early theoretical studies \citep{1978ApJ...222..604P,1980PASJ...32..473N,1982ApJ...257..752F,1982ApJ...261..649S}, and was confirmed through extensive parameter studies \citep{1985STIN...8617240S,1986ApJ...310..222P,1986ApJ...308..721T,1995ApJ...445..789P,1995ApJ...448..807P}. Modern multidimensional simulations and observations continue to refine these relationships \citep{2005ApJ...621L..53S,2015MNRAS.446.1924H,2019ApJ...879L...5H,2021MNRAS.501..201H,2024MNRAS.527.4806V}, but the fundamental framework established in the 1980s and 1990s remains valid \citep{1992ApJ...393..516L,2004ApJ...600..390T}.

The mass of the WD is a critical parameter, with more massive WDs experiencing more violent and frequent eruptions \citep{1972ApJ...176..169S,1978A&A....62..339P,1978ApJ...222..604P,1985STIN...8617240S,1986ApJ...310..222P,1995ApJ...445..789P,1997ApJ...477..356K,2005ApJ...623..398Y,1992ApJ...393..516L,2024MNRAS.527.4806V}.
The evolution models of classical nova eruptions predict that less massive white dwarfs demonstrate more drastic changes in their light curves compared to more massive ones \citep{1986ApJ...310..222P,1995ApJ...445..789P,1997ApJ...477..356K,2005ApJ...623..398Y,2013ApJ...777..136W,Hillman_2016}. This occurs because lower-mass WDs require larger ignition masses and reach higher peak temperatures, leading to more violent eruptions with longer-lasting nuclear burning and more complex evolution \citep{1986ApJ...303L...5S,1986ApJ...308..721T,1995ApJ...448..807P,2004ApJ...600..390T}.
Higher mass WDs have stronger surface gravity, leading to higher pressures at the base of the accreted envelope, which results in higher peak burning temperatures, shorter recurrence times, and faster evolution \citep{1986ApJ...308..721T,2004ApJ...600..390T}.

According to \cite{2022MNRAS.511.5570H}, the presence of heavy elements in the accreted envelope also significantly influences the eruption dynamics. Heavy elements, particularly CNO isotopes, act as catalysts in the hydrogen burning via the CNO cycle, dramatically increasing the rate of energy generation at a given temperature \citep{1972ApJ...176..169S,1974ApJS...28..247S,1976ApJ...208..819S,Starrfield_2016}. This enhancement results in lower ignition masses, shorter accretion phases, and more violent eruptions with higher ejection velocities \citep{1998MNRAS.296..502S, 2000ApJS..127..485S}. The metallicity of the accreted material can vary the ignition mass by factors of 2-3 and the recurrence time by an order of magnitude for the same WD mass and accretion rate \citep{1985ApJ...291..812K,1998ApJ...494..680J,starrfield2012theoreticalstudiesaccretionmatter}.

The resulting eruption can temporarily reach about 50 thousand times the solar luminosity, outshining its companion \citep{2023JHEAp..38...22B}. Despite their prominence, the detection and comprehensive understanding of novae continue to pose challenges for astronomers. These challenges include accurately modeling the complex interplay of accretion, thermonuclear runaway, and mass ejection, as well as understanding the long-term evolution of nova systems \citep{2012MNRAS.424L..69E, 2021MNRAS.505.3260H}.

\subsection{Nova light curves}
Light curves of astrophysical objects are a fundamental tool in astrophysics for studying a wide range of celestial objects and phenomena. Analyzing the patterns and features in the light curves allows us to gain insights into the physical properties, behavior, and underlying mechanisms of these objects.

The light curve and spectral evolution of a classical nova are direct consequences of the physical state and expansion of the TNR heated envelope. The eruption begins with a violent TNR on the white dwarf’s surface in an accreted, degenerate hydrogen envelope, often mixed with WD material, which drives rapid energy release and envelope expansion \citep{2007JPhG...34..431J,2021ARA&A..59..391C}. During the runaway, temperatures reach an order of $10^8$ K; theory predicts a short-lived early ultraviolet/soft X-ray flash as the effective temperature rises prior to substantial expansion \cite[]{2014MNRAS.437.1962H}, although this phase is difficult to catch observationally \citep{2021ARA&A..59..391C}. To shed the energy and excess mass, the envelope undergoes a prolonged expansion; as it expands, it cools, shifting the spectral peak from the ultraviolet into the optical and producing the observed rapid rise to optical maximum, with the bolometric output remaining near Eddington for some time \citep{2005ASPC..330..265H,2007JPhG...34..431J}. Following the ejection of much of the envelope, the shell contracts and heats, often entering a super-soft phase \cite[]{Orio2001,Ness2011,Page2015}, and finally relaxes into a gradual optical decline  \citep{1986ApJ...310..222P,2007JPhG...34..431J,2014MNRAS.437.1962H}.

Fast novae are generally observed to be dominated by optically thin free-free emission in their optical and near-infrared (NIR) spectra during the early-to-mid decline phase (with the strongest free–free contribution typically between days to a few weeks post-peak). This leads to a rapid decline in brightness that is only weakly dependent on the chemical composition of the ejecta or the exact white dwarf (WD) mass. \citep{1981PASP...93..165D,1989AJ.....97.1622C,1995ApJ...452..704D,2006ApJS..167...59H,2014ASPC..490..327M}
In a rough but fairly effective way, the energetics of a nova explosion is parametrized by the time $t_2$ (or $t_3$) that the nova's visual light curve takes to decline by 2 (or 3 magnitudes) from maximum \citep{1943POMic...8..149M,1957gano.book.....G,1981PASP...93..165D}.
Faster novae, with shorter $t_2$ have larger bolometric luminosities and higher ejection velocities. However, shock processes within nova ejecta, observed through gamma-ray emission, also contribute to the bolometric luminosity during an eruption, influencing the observable decline characteristics \citep{2020NatAs...4..776A}.
In summary, optical and IR light curves often have different shapes; the optical decays as the ejecta expands and cools, while the IR stays bright due to free-free emission from the ionized gas.

However, the extremely erratic nature of some nova light curves, which exhibit halts, prolonged, fluctuating maxima or extended plateaus, makes it difficult to determine and define the exact timing of the peak brightness and consequentially, the corresponding $t_2$ \citep{2010AJ....140...34S,2011ApJ...735...94K,Cao_2012,Eyres2017,aydi2020directevidenceshockpoweredoptical,Mason2020,2022MNRAS.515.1404H,Lightcurve}.

As the ejecta expands and the density decreases, the nova transitions to an emission-dominated spectrum. Observing this transition phase provides insight into the chemical composition, velocity structure, and evolution of the ejecta, which are important to understand the processes occurring during and after the eruption \citep{1993fain.proc..127W,Shore_2013}. The study of novae also provides insights into the chemical evolution of galaxies, as they contribute to the enrichment of the interstellar medium with newly synthesized elements \citep{1992A&A...266..232D,1994A&A...286..786D,2012ApJ...755...37H,2019NatAs...3..725L,1995ApJ...452..704D,1999A&A...352..117R,2015ApJ...808L..14I,Starrfield_2016,2023A&A...679A..72M}

The behavior of novae in both the optical and infrared bands provides valuable insight into their physical properties.
Faster declining novae are intrinsically more infrared bright relative to their optical emission because fast novae are typically associated with more massive white dwarfs and more powerful explosions, which result in ejecta that expand and cool more rapidly. 
In contrast, slower novae, which lower mass white dwarfs generally power, do not cool as quickly and have a greater contribution from photospheric emission, so their spectral energy distributions (SEDs) remain relatively bluer at peak \citep{2014ApJ...785...97H,2015ApJ...798...76H}. This strong anti-correlation supports the idea that the SED of fast novae undergoes a significant shift toward longer (IR) wavelengths at peak brightness. This spectral shift is consistent with the WD's envelope expanding after the TNR and the onset of mass loss, leading to rapid cooling. 

The onset of dust formation, while typically occurs a few tens of days after optical peak, is not uniform across all novae. In some fast novae such as V5579 Sgr, early signs of dust formation can appear as soon as 15 to 20 days after the outburst, detected through near-infrared photometry showing excess emission in the J (1.25 $\mu$m), H (1.65 $\mu$m), and particularly K (2.2 $\mu$m) bands, with the IR excess becoming detectable during that interval and evolving noticeably between 20 and 70 days \citep{2024PASA...41...51R}. 
Even in these cases, the major contribution of dust to the overall emission often becomes prominent after about 60 days, particularly as the dust mass builds up and the ejecta expands further, with substantial dust formation occurring at later epochs \citep[typically 60 to 90 days or even beyond,][]{2017gacv.workE..64K,2025MNRAS.536.2661B}. 

\subsection{Multi-band observations of novae}
A wide effort to observe novae across the electromagnetic spectrum has been put forth to understand the physics of nova eruptions, yielding key discoveries regarding the mass ejection mechanisms and shocks contributing to their luminosity \citep{2021ARA&A..59..391C}. Numerous extra-Galactic surveys in optical and UV bands have been carried out \cite[e.g., ][]{Mandel2023,Shara2023} finding links between these band that contribute to the understanding of light curve features. 
Multi-band observations of novae are essential for disentangling the diverse physical mechanisms at play, with each spectral band providing unique diagnostic insights \citep{2017ApJ...840..110W, 2023JHEAp..38...22B}. In particular, the peak of the emission (approximated by blackbody radiation) shifts with the nova’s temperature. Early in an eruption, the photosphere is very hot, and the peak emission can lie in the ultraviolet (UV) or optical bands; later, as the ejecta cools and expands, IR and even radio emission become important. Thus, multi-band photometry 
provides a more complete picture of the nova’s evolution.
IR observations of novae, particularly longward of $3~\mu$m, provide crucial insights into the nova phenomenon beyond what optical observations alone can achieve. The IR emission exhibits a characteristic sharp rise in flux that coincides with a deep decline in the visual light curve.

X-ray and gamma-ray observations, such as those from Swift, Chandra, or Fermi, have revealed hot, shocked gas and even particle acceleration in some novae \citep{Mukai2008,2009AJ....137.4160N,Ackermann2014,2014Natur.514..339C,2015MNRAS.450.2739M,2020NatAs...4..776A}. The X-ray emission arises from shocks internal to the ejecta or from the interaction between the ejecta and the companion wind, reaching temperatures of $10^6-10^8$ K \citep{2006Natur.442..279O,2006ApJ...652..629B,2010MNRAS.401..121P,2011ApJ...727..124O}. The unexpected discovery of GeV gamma-ray emission from several novae by Fermi-LAT demonstrates that shocks can accelerate particles to relativistic energies, fundamentally changing our understanding of nova physics \citep{2010Sci...329..817A,Ackermann2014,2016ApJ...826..142C,2018A&A...609A.120F,2020NatAs...4..776A}. In this context, novae could also represent unexpected sources of high-energy neutrinos, as recently suggested by \cite{2023JCAP...03..015G}, further supporting the idea that shock-powered novae can act as cosmic accelerators. Therefore, combining data from UV through optical to IR allows us to follow the nova from its hottest, luminous phase into its dust-cooling phase, filling in the gaps that single-band observations leave \citep{2011ApJS..197...31S,Shore_2016,2016MNRAS.462.1591E,2017MNRAS.469.4341M,2022MNRAS.514.2239S}.

Optical and UV observations capture the hot, early-phase continuum and strong emission lines of highly ionized species, while IR observations are sensitive to warm dust and recombination lines that appear as the ejecta cools. Optical colors, such as $(B-V)$ or $(B-I)$, trace the temperature and reddening of the ejecta. It is also known that novae become redder as they fade. Studies have found that the intrinsic $(B-V)_0$ color of novae is fairly constant near peak and after decline. \cite{2025MNRAS.538.2339C} report an average intrinsic $(B-V)_0\approx+0.20$ at peak and $(B-V)_0\approx -0.03$, two magnitudes below peak. Similarly, \cite{1987A&AS...70..125V} found $(B-V)_0 \approx -0.02\pm0.04$ at $t_2$. Such behavior allows novae to be used as "standard crayons" for estimating interstellar reddening from photometry alone \citep{2025MNRAS.538.2339C}.

The color evolution of novae has been extensively studied in the optical regime, revealing systematic patterns that reflect the underlying physics. \cite{1989AJ.....97.1622C} established that optical colors follow predictable tracks during decline, with faster novae showing different color evolution than slower ones. \cite{2010AJ....140...34S} analyzed 93 nova light curves and found that optical color indices correlate with speed class, while \cite{2014ApJ...785...97H} demonstrated that 
$(U-B)$ and $(B-V)$ colors can distinguish nova subtypes. More recently, \cite{2017MNRAS.469.4341M} showed that novae emitting gamma rays display distinct optical color evolution compared to non-gamma-ray novae.

Although optical colors have been well characterized and UV-optical studies have revealed ionization evolution, the connection between optical and mid-IR bands has received limited attention. Previous IR studies have typically focused on individual objects or late-time dust formation rather than the color evolution of the nova eruption at early times. The Wide-field Infrared Survey Explorer (WISE) mission, with its all-sky coverage at 3.4 $\mu$m (W1) and 4.6 $\mu$m (W2), provides an unprecedented opportunity to study optical-mid-IR colors for a large sample of novae uniformly.
 
\section{Method - data extraction from the WISE archive} \label{sec:style}

The WISE satellite, launched by NASA in December 2009, has imaged the entire sky in four mid-IR bands (3.4, 4.6, 12, and 22$\mu$m) \citep{2010AJ....140.1868W}. WISE surveyed the sky with sensitivities far exceeding those of previous IR missions and cataloged approximately 750 million sources.

With its all-sky coverage and sensitivity to infrared emissions, WISE data are an ideal resource for studying transient and variable objects.

WISE observed each sky region multiple times per survey cycle, typically 10 -12 exposures per source over $\sim$1 to 2 days, with a cadence that allowed detection of both short-term and long-term variability.
The WISE database provides over 12 years of infrared monitoring, during which it captured multi-epoch observations of celestial sources across the entire sky approximately every six months.
We accessed WISE time-domain data via NASA’s Infrared Science Archive (IRSA). 

\sloppy
Using the \textsc{AllWISE} and \textsc{NEO\-WISE-Reac\-ti\-va\-tion (NEOWISE-R)} Source Catalogs 
and Single Exposure (L1b) databases\footnote{\footnotesize 
The \textsc{AllWISE} Source Catalog is a comprehensive infrared sky survey catalog released by NASA using data from the Wide-field Infrared Survey Explorer (WISE) mission, including the original WISE cryogenic mission (2010) and the NEOWISE Post-Cryo phase. The \textsc{NEO\-WISE-Reac\-ti\-va\-tion (NEOWISE-R)} Source Catalog contains data collected after WISE was reactivated in late 2013, primarily to hunt for near-Earth objects (NEOs), but also useful for time-domain infrared astrophysics. Individual time-stamped WISE/NEOWISE images and detections from a single pass over a region of the sky typically contain $\sim$10--12 observations over a few days.} We queried infrared photometry of all sources based on their light curves. We extracted time-resolved data primarily in the $W1$ and $W2$ bands (3.4 and 4.6~$\mu$m, respectively) due to their availability throughout the entire duration of the mission.

 We constructed an algorithm to filter the entire WISE database to identify transient sources and retrieve their light curves. Our computational algorithm identified many IR sources using the following selection criteria:
\begin{enumerate}[label=(\roman*), topsep=5pt]
    \item 
Variability: We require that the light curve show a significant variability threshold. We accomplish this by treating short and long-duration events differently. For short-duration events ($<$20 days), we require the $W1$ and $W2$ variability to be greater than $\times3$ the mean uncertainty (3$\sigma$). 
For long-duration trends ($>$20 days) we require the variability of $W1$ and $W2$ to be $>0.5\rm mag$ per year.
These thresholds ensure changes are not due to noise or typical stellar variability.
\item
Consistency: We require a robust $W1-W2$ Pearson correlation coefficient to exceed 0.5 to ensure that the source does not vary randomly or due to noise, and that the correlation is not driven by a single outlier.
\item
Sampling: We require that the light curve (LC) contain at least 10 valid (non-NaN) observations after filtering, to enable the computation of reliable trends and correlations. This safeguard helps to eliminate both masquerading sources (interlopers) and intrinsically slow-changing background objects, leaving us with a more reliable pool of genuine candidates.
\item
Location: We require positional consistency by allowing detections only within 5 standard deviations of the median RA and DEC. This is to eliminate blended sources and to avoid confusion with nearby sources.
\item 
Temporal trend: We require a strong global correlation or a partial post-peak correlation of $W1$ and $W2$ with time. Our criteria for this, is that $W1$ (and $W2$) correlates with time with a Pearson coefficient of at least 0.5, if not global, then at least a partial segment of the LC after the peak shows this behavior. 
This is to catch events where the entire lightcurve may not be monotonic (rises, peaks, then fades), but at least part of it (after the peak) shows consistent brightening or fading.

\item 
In addition to the above criteria, we require a consistent rising or fading trend in both bands by requiring that both \textsc{PRw1mjd} and \textsc{PRw2mjd} \footnote{\textsc{PRw1mjd} and \textsc{PRw2mjd} are the Pearson correlation coefficients between the Modified Julian Date (MJD) and the W1 or W2 magnitudes, respectively} are $>0.5$ (or $<-0.5$).

This ensures that regardless of whether the trend came from the entire curve or just a segment, both bands must show strong and consistent correlation over time.
\end{enumerate}

By rigorously applying our well-defined selection criteria, we were able to pull a total of 1,900 targets from this catalog.

\begin{table*}[ht!]
\centering
\caption{Summary of optically confirmed novae that our pipeline successfully identified through a thorough examination of the WISE catalog. For each source, key parameters are provided, and these parameters serve both to confirm the identity of the sources and to provide context for subsequent analysis of their infrared behavior. The table highlights the effectiveness of our pipeline in filtering novae across multiple epochs of WISE observations. \citep{mukai_novae} \citep{aavso_vsx}.}
\label{tab:astro_table}
\resizebox{\textwidth}{!}{%
\begin{tabular}{cccccc}
\hline\hline
S/N & RA (deg) & Dec (deg) & ID & Eruption Year & Reference \\
\hline
    1  & 208.3650342
   & -67 25 00.9  & FM Cir                       & 2018 & \cite{2018PZ.....38....5K} 
   \\ 
   2 & 283.1457076
  & -0.3118109
  & V1724 Aql                    & 2019 & \cite{2012CBET.3273....2M, 2012CBET.3273....1N}
  \\ 
    3 & 159.0641998
   & -59.5982604
  & V906 Car                     & 2018 & \cite{2018ATel11460....1L, 2018ATel11456....1S}
  \\ 
    4 & 208.688903
  & -59.1511814
  & V1369 Cen                    & 2014 & \cite{ATel5621, 2016ApJ...826..142C} 
  \\ 
    5 & 200.2305555
  & -63.70534935
 & V1405 Cen                    & 2021 & \cite{2017ATel10387....1S, Novae} 
 \\ 
    6 & 347.0196151
 & 60.7810973
   & V809 Cep                     & 2013 & \cite{CBAT_V2659Cyg} \\ 
    7 & 313.5989652
   & 60.2852282
   & V962 Cep                     & 2014 & \cite{IAUC9270} \\ 
   8 & 108.4410002
  & -21.2086995
  & V435 CMa                    & 2018 & \cite{2018ATel11475....1S, 2021ApJ...910..120K}
  \\ 
    9 & 305.426457
  & 31.0581885
   & V2659 Cyq                     & 2014 & \cite{2018yCat.1345....0G}\\ 
    10 & 305.8778647
  & 20.7676814
   & V339 Del                     & 2013 & \cite{2013ATel.5282....1S, 2018yCat.1345....0G}
   \\ 
    11  & 251.2092771
  & -45.2632242
  & V1674 Her                   & 2021 & \cite{2021ATel14746....1W, 2021ATel14835....1S} \\ 
    12 & 232.2573736
  & -44.8277523
  & V407 Lup                     & 2016 & \cite{2016ATel.9539....1S, 2018MNRAS.480..572A} 
  \\ 
   13 & 201.130338
  & -72.17506835
 & V0415 Mus                  & 2022 & \cite{CBET5135} \\ 
    14 & 262.3060403
   & -18.77065325
 & V2944 Oph                    & 2015 & \cite{2015ATel.7367....1M}\\ 
 15 & 259.6874851
  & -24.9062141
  & V3663 Oph        & 2017 & \cite{2017ATel10959....1C} \\  
    16 & 259.5265429
  & -32.0743186  & V1661 Sco                    & 2018 & \cite{2018ATel11209....1S}\\ 
    17 & 252.2069734
  & -44.95091155
 & V1662 Sco                    & 2018 & \cite{2018ATel11289....1S}\\ 
    18 & 255.9479729
   & -38.28260045
  & V1663 Sco                    & 2018 & \cite{2018ATel11348....1S} 
  \\ 
       19  & 142.4711788
  & -56.2905797
   & V6567 Sco                 & 2022 & \cite{ATel15395} \\ 
   20 & 279.9987693
  & -10.4282745
  & V659 Sct                     & 2019 & \cite{cbet4690}\\ 
  21 & 272.26431105 & -11.2095807 & V0556 Ser & 2013 & \cite{2015PZ.....35....3K} \\
    22 & 276.2865037
  & -22.6008877
  & V5666 Sgr                    & 2014 & \cite{2021ApJ...910..120K}\\ 
    23 & 273.6048596
  & -25.9096905
  & V5667 Sgr                    & 2015 & \cite{astronet_varstars} \\ 
    24 & 279.2368521
  & -28.9277277
 & V5668 Sgr                    & 2015 & \cite{astronet_varstars} \\ 
    25 & 270.8863928
   & -28.2683858
  & V5669 Sgr                    & 2015 & \cite{2015IAUC.9277....3F, 2015ATel.8101....1W}\\ 
    26 & 275.2176926
  & -28.370068
  & V5856 Sgr                    & 2016 & \cite{2016ATel.9678....1L}\\ 
     27 & 271.0393662
  & -18.06548645
 & V5857 Sgr        & 2018 & \cite{cbet4504}\\ 
  28 & 268.834371
   & -23.3984378
 & V5862 Sgr            & 2014 & \cite{2019IBVS.6261....1K}\\ 
     29 & 169.2797087
  & -64.6158061
   & V5927 Sgr                    & 2014 & \cite{2016ATel.9215....1M} \\ 
    30 & 223.7802765
  & -60.4445083
  & V5928 Sgr                    & 2014 & \cite{2013AcA....63...21S, 2022ApJS..260...46I}\\ 
  31 & 270.49327625 & -26.0573552 & V5933 Sgr & 2014 & \cite{2019PZ.....39....3K} \\
  32 & 268.7499736
  & -21.3778449
  & V6593 Sgr       & 2020 & \cite{2020ATel14064....1A, 2020ATel14062....1D} \\ 
    33  & 282.2711411
  & -19.03454245
 & V6594 Sgr                 & 2021 & \cite{2021ATel14488....1M, 2021ATel14487....1S} \\ 
    34 & 59.6231632
  & -54.77810465
 & YZ Ret                       & 2020 & \cite{2020ATel13867....1A, 2021RNAAS...5..150S} \\
 35  & 261.5807359
  & -33.45298175
 & ASASSN-21pa .                & 2021 & \cite{2021ATel14880....1A} \\ 
  36  & 270.1870315
   & -21.66128615
 & AT 2021aadi                  & 2021 & \cite{2021ATel14950....1D} \\ 
 37  & 285.8123314
   & 1.3411823
    & NOVA Aql 2019               & 2019 & \cite{2021ATel14657....1D} \\  
    38 & 80.616083
  & -69.6091309
  & PNV J05222788-6936333      & 2022 & \cite{2022ATel15392....1A}\\ 
 39 & 89.4930282
  & -74.9025
     & LMCN 2013‑10a             & 2013 & \cite{2014AcA....64..197W, 2016ApJS..222....9M}\\ 
  40  & 82.5700055
  & -73.2697622
  & LMCN 2017-11a    & 2017 & \cite{2018ATel11132....1C, 2019arXiv190309232A} \\   
  41  & 85.4440036
    & -71.8104442
  & LMCN 2018-07a                 & 2018 & \cite{2018ATel11959....1A, 2018ATel11863....1P} \\ 
\hline
\end{tabular}
}
  
    \tablecomments{ \textit{
$^{1}$ LMCN stands for Large Magellanic Cloud Nova. It is a standardized naming convention for novae occurring in the Large Magellanic Cloud (LMC).\\
$^{a}$ LMCN 2017-11a is also known as ASASSN-17pf 
$^{b}$ ASASSN-21pa is also known as Gaia21dyi 
$^{c}$ AT 2021aadi is also known as Gaia 21ejq and PGIR21git. 
$^{d}$ LMCN 2018-07a is also known as Gaia18bvg, ASASSN-18pf and AT 2018dya   \citep{2018TNSTR.983....1S}.
$^{e}$ NOVA Aql 2019 is  also known as Gaia19buy, PGIR20fbe, ZTF19aaontsz \citep{2021ATel14657....1D}.
$^{f}$ LMCN 2013‑10a is also known as OGLE-2013-NOVA-02. 
$^{g}$ PNV J05222788-6936333 is also known as OGLE-LMC-SC6 418248.
$^{h}$ V6567 Sco is also known as Gaia22alz and AT2022bpq.
$^{i}$ V5856 Sgr is also known as ASASSN-16ma}}

 \label{tab:astronomical_objects}
\end{table*}

\section{Results}
\subsection{WISE catalog}
We cross-matched approximately 1,900 algorithmically selected WISE candidates with SIMBAD\footnote{SIMBAD: Set of Identifications, Measurements, and Bibliography for Astronomical Data --- an astronomical database of objects beyond the Solar System} database 
and found 101 WISE objects that have a previously confirmed optical observation.
Of these 101 objects 41 sources are previously confirmed novae, 14 sources are cataloged as unclassified transients or variable stars, and 46 are listed under other categories such as Young Stellar Objects, Blazars, Long Period Variables, Orion Variables, or mid-IR sources. The high rate of sources pulled from our initial search reflects the broad sensitivity of our algorithm, which was intentionally tuned to favor recall over precision. With further manual vetting, we confirmed these events and excluded artifacts and low-SNR\footnote{SNR: Signal-to-Noise Ratio} detections.

Using the light curve characteristics of the confirmed novae recovered by our pipeline as a reference, we performed a visual analysis of the light curves of the remaining $\sim1,800$ unmatched sources. This process yielded a catalog of 58 high-confidence objects exhibiting strong nova 
behavior. Combining these results, we obtained a catalog of 113 sources of interest: 41 optically confirmed novae and 72 candidates (14 optically known but unclassified variable sources and 58 previously unreported sources in the optical). 

The properties of the 41 known V-band novae for which we also have mid-IR (W1) band LCs are summarized in Table \ref{tab:astronomical_objects}. Figures  \ref{fig:V-W1A} and \ref{fig:V-W1} display the V-band light curves obtained from the American Association of Variable Star Observers (AAVSO) alongside the corresponding W1 band light curves from WISE, showing the subset of novae where the V and W1 observations overlap temporally. For these cases, the $V-W1$ color evolution was calculated from the overlapping data and is plotted in green.
Figures \ref{fig:V-W11A} and \ref{fig:V-W11} present the V and W1 light curves for novae that lack temporal overlap between the two bands; therefore, $V-W1$ could not be determined in those cases.

\subsection{ V-W1 model application}
Studies of dusty novae in particular indicate that dust formation commonly starts in the interval between about 20 and 40 days after eruption \citep{2008clno.book..167G, 2014MNRAS.440.3402M}. 
While incipient dust formation may begin in such an early phase, its overall impact on the spectral energy distribution, and hence on the $V-W1$ color, is modest until the dust mass grows to a level that produces significant optical absorption and infrared re-emission. 
In moderately fast novae, such as V5584 Sgr, dust condensation is noted around 45 ± 5 days after maximum brightness, while slower or dust-rich systems may not display substantial dust effects until closer to 60 or even 100 days after eruption \citep{2018A&A...619A.104S, 2025MNRAS.536.2661B}. 
In general, significant dust formation in classical novae typically begins 60 - 100 days after optical maximum, when the ejecta temperature drops below the condensation temperature ($\sim1500-2000 K$) for graphitic or silicate grains \citep{1988ARA&A..26..377G,1997MNRAS.292..192E,2003AJ....125.1507S,2017MNRAS.469.1314D,Gehrz_2018}. 
Since dust formation can contribute to mid-IR excess, we restricted our analysis to the first 60 days after $V-W1$ peak, before the typical onset of significant dust formation, so that the observed trend reflects intrinsic emission processes rather than dust effects.

Figure \ref{fig:MagVsTime} presents the natural evolution of the $V-W1$ color of the 25 confirmed novae 
for which we were able to produce $V-W1$ color curves. This figure shows their natural evolution without aligning their peaks, providing an initial overview of their diverse photometric behaviors. This preliminary visualization revealed broadly similar  early decline trends in many systems, which motivated a more detailed examination using a fixed peak reference. Based on this, Figure \ref{fig:Declineplot} illustrates the comprehensive color evolution of $V-W1$ for the 25 confirmed novae. 
capturing both early and late evolution phases with a common peak to better assess their rates of decline.

 To emphasize the clean and dust-free evolution period, Figure \ref{fig:Declineplot2} narrows down to the first 60 days post-maximum. 
Throughout this early period, the $V-W1$ color curves demonstrate a strikingly linear decline across multiple systems. 
The synchronization in these color curves suggests a common underlying mechanism governing intrinsic infrared emission of novae before significant dust formation. This observed linearity indicates that the early $V-W1$ color evolution is a reliable indicator of the behavior of novae, establishing a solid baseline for identifying subsequent deviations driven by dust. 
\\
Also, computing the difference between the $V$ and $W1$ brightnesses, effectively removes the influence of distance because the apparent magnitude of a source depends on its absolute magnitude and distance through the distance modulus. 
Therefore, our $V-W1$ color curves do not require distance calibration, which makes this method robust and efficient for comparing nova behavior independent of distance uncertainties.

\begin{figure*}[t!]
\centering
\begin{tabular}{ccc}
\includegraphics[width=0.30\textwidth]{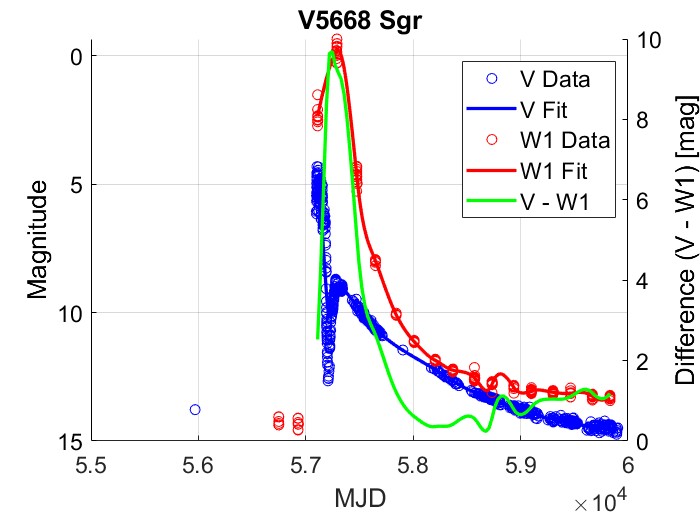} &
\includegraphics[width=0.30\textwidth]{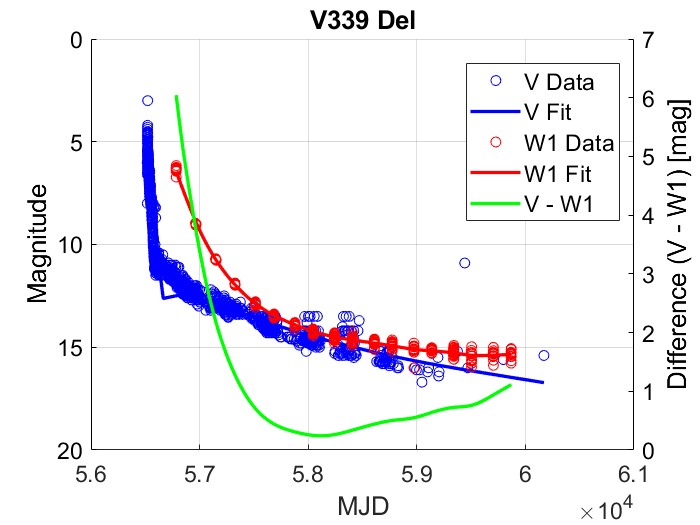} &
\includegraphics[width=0.30\textwidth]{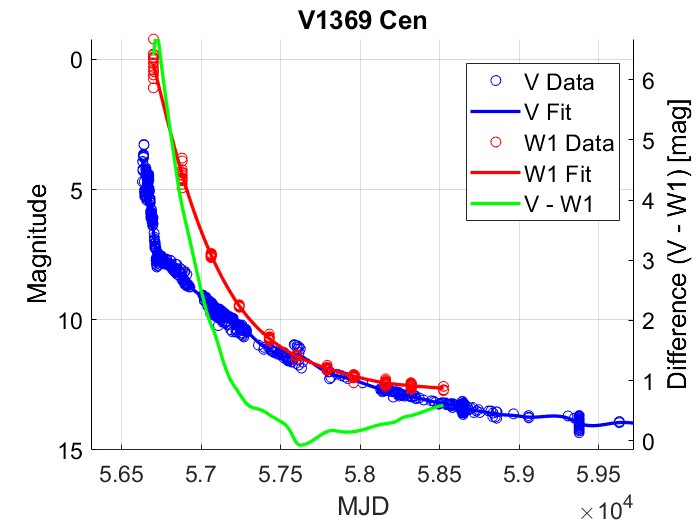} \\

\includegraphics[width=0.30\textwidth]{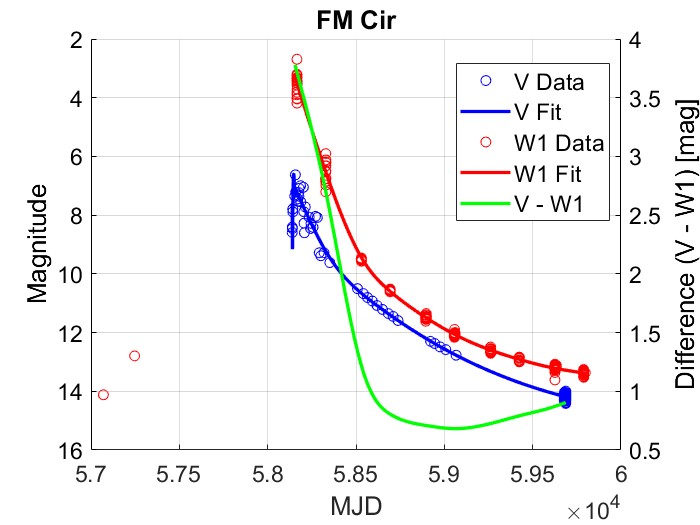} &
\includegraphics[width=0.30\textwidth]{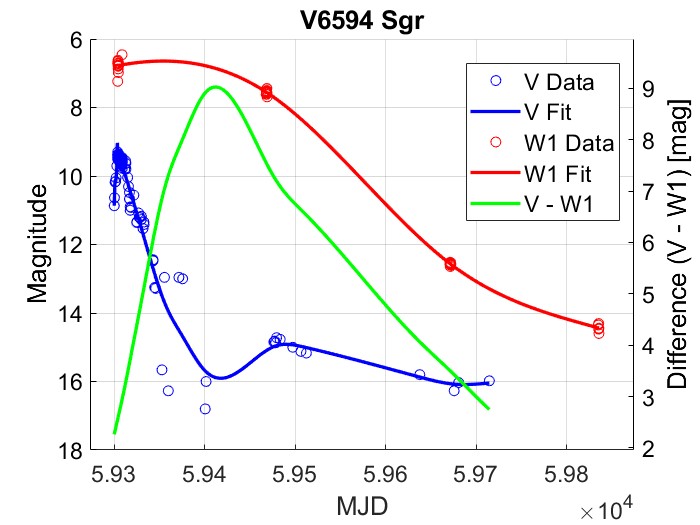} &
\includegraphics[width=0.30\textwidth]{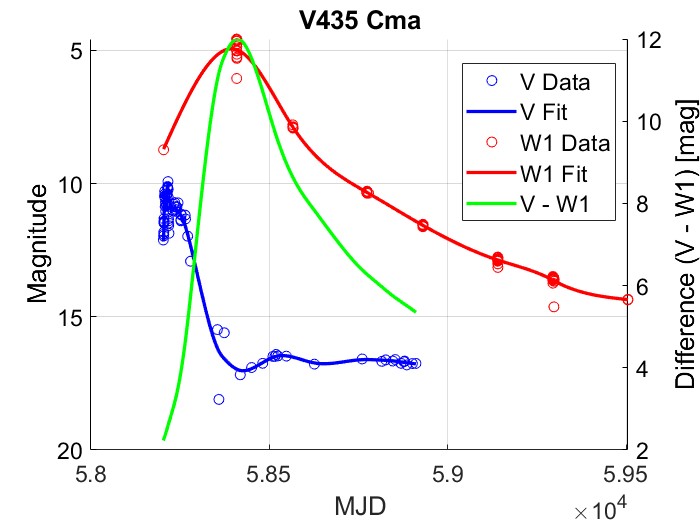} \\

\includegraphics[width=0.30\textwidth]{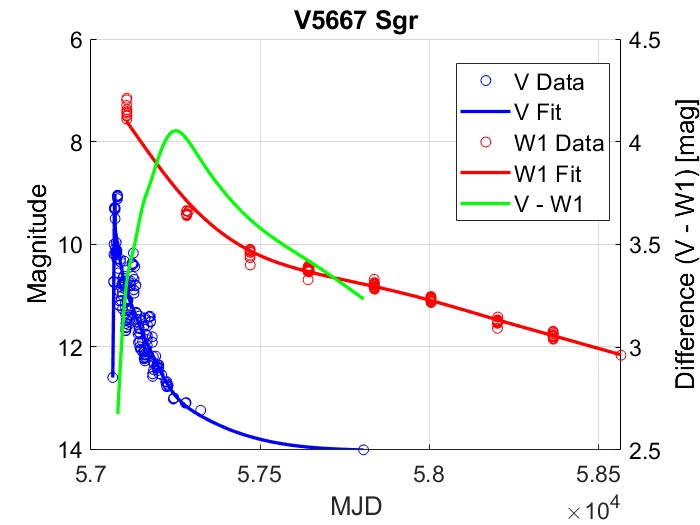} &
\includegraphics[width=0.30\textwidth]{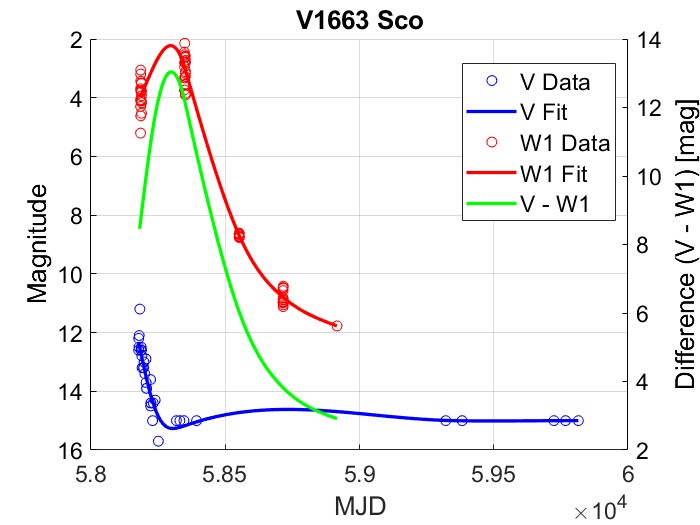} &
\includegraphics[width=0.30\textwidth]{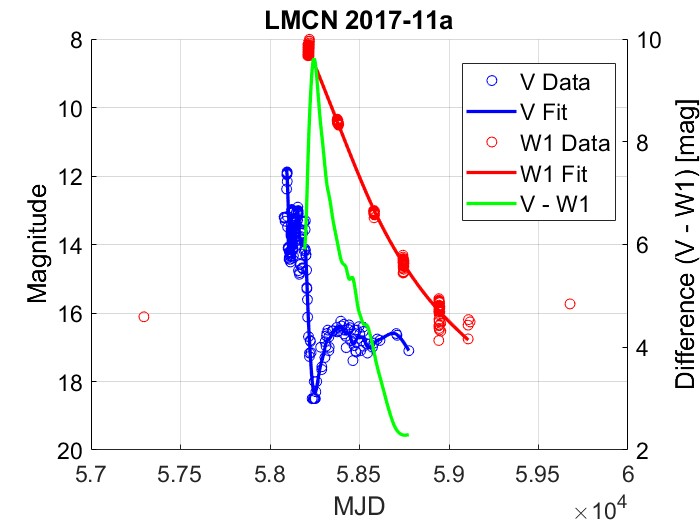} \\

\includegraphics[width=0.30\textwidth]{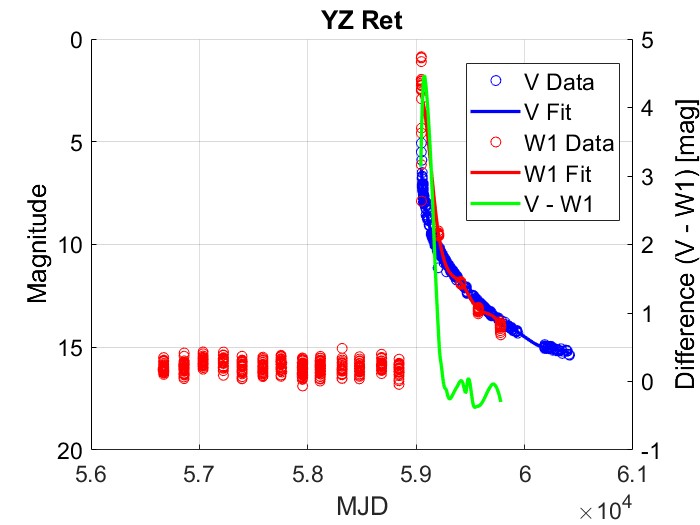} &
\includegraphics[width=0.30\textwidth]{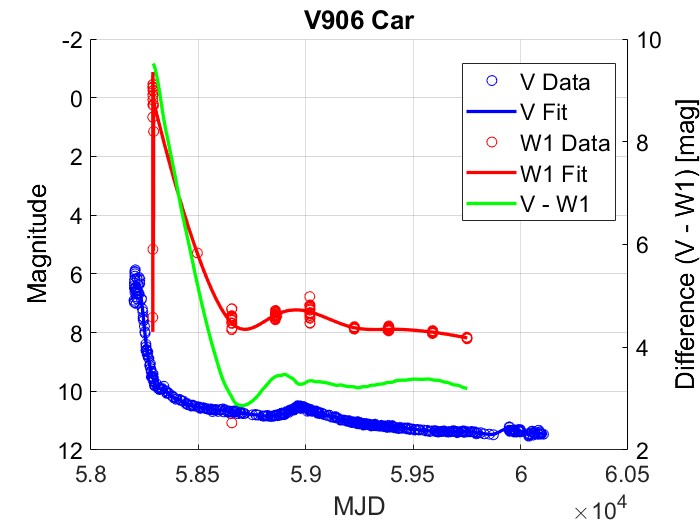} &
\includegraphics[width=0.30\textwidth]{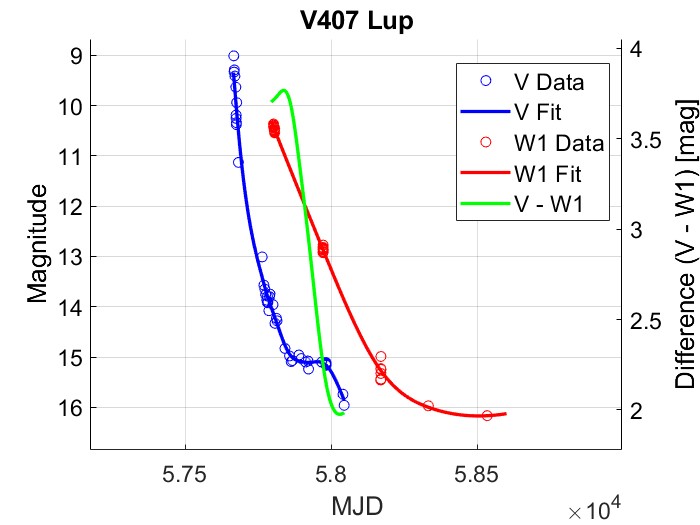} \\

\includegraphics[width=0.30\textwidth]{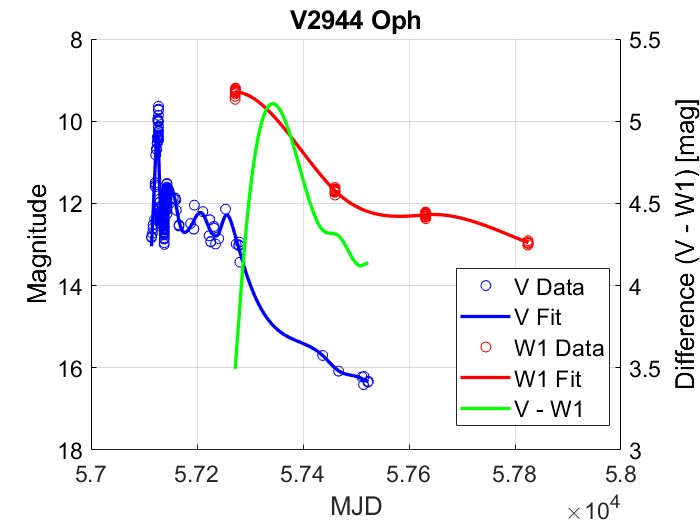} &
\includegraphics[width=0.30\textwidth]{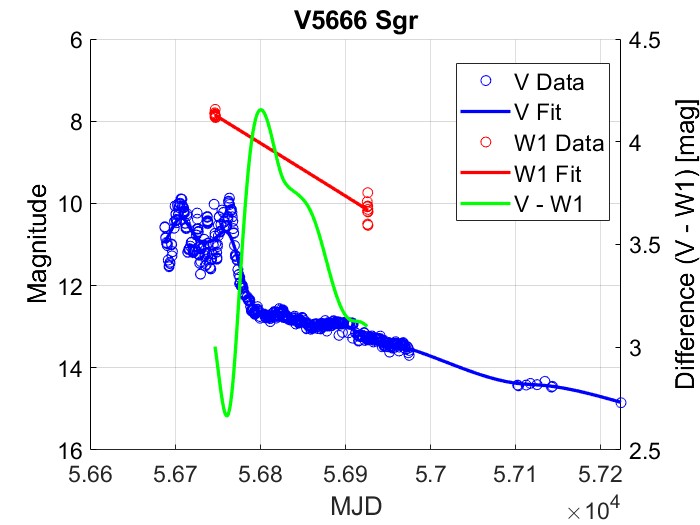} &
\includegraphics[width=0.30\textwidth]{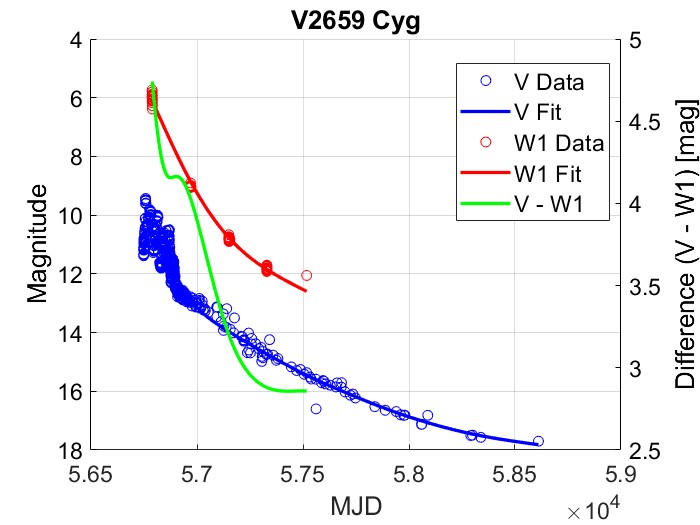} \\
\end{tabular}
\caption{$V-W1$ color difference model applied to our identified samples of known novae. Individual data points are shown in red (WISE W1 band) and blue (optical V band), capturing the temporal evolution of each emission component. The green curve represents the computed ($V-W1$) color difference, which serves as a diagnostic of the nova’s emission characteristics.}
 \label{fig:V-W1A}
\end{figure*}

\begin{figure*}[t!]

\centering
\begin{tabular}{@{}ccc@{}}
\includegraphics[width=0.32\textwidth]{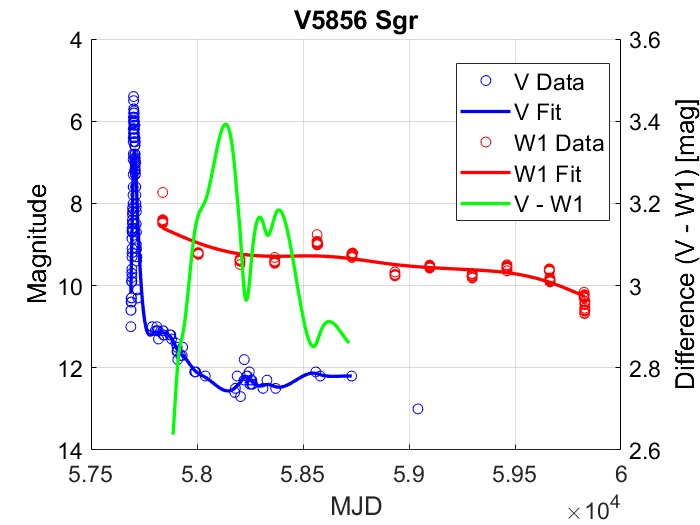} &
\includegraphics[width=0.32\textwidth]{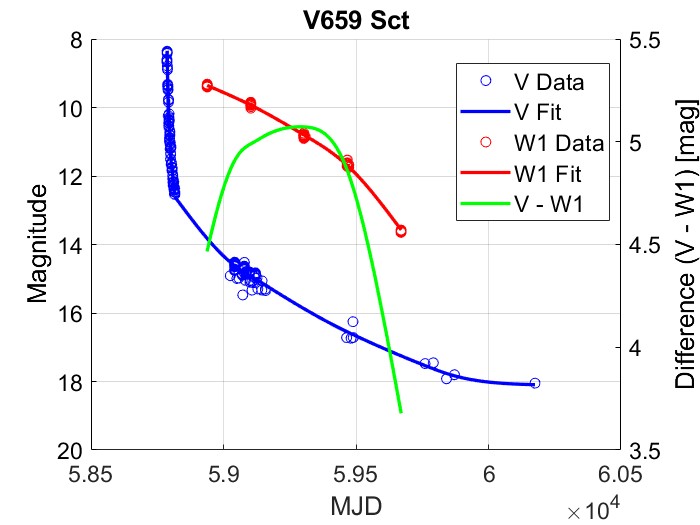} &
\includegraphics[width=0.32\textwidth]{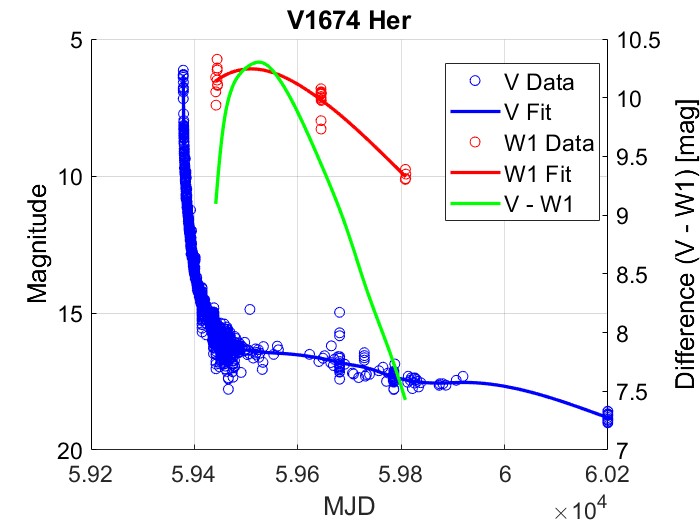} \\

\includegraphics[width=0.32\textwidth]{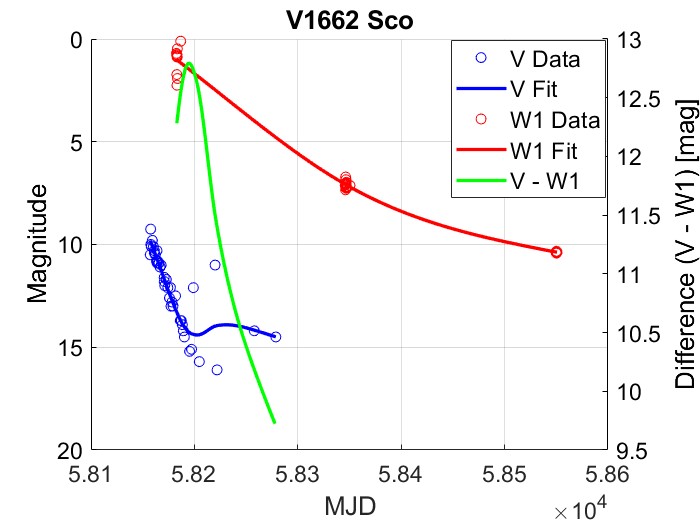} &
\includegraphics[width=0.32\textwidth]{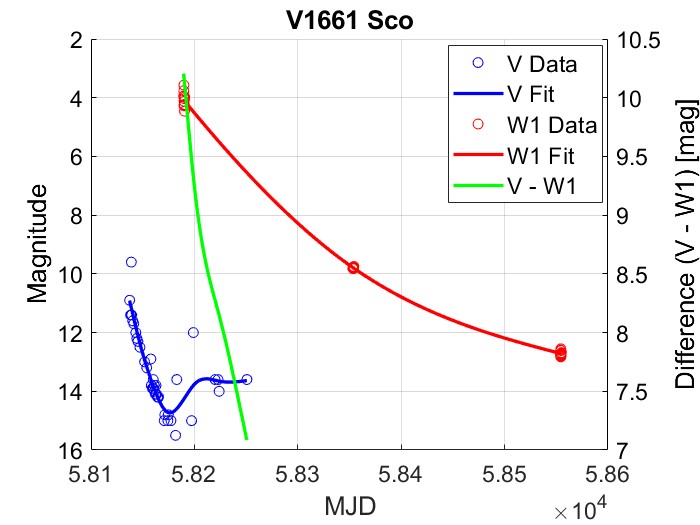} &
\includegraphics[width=0.32\textwidth]{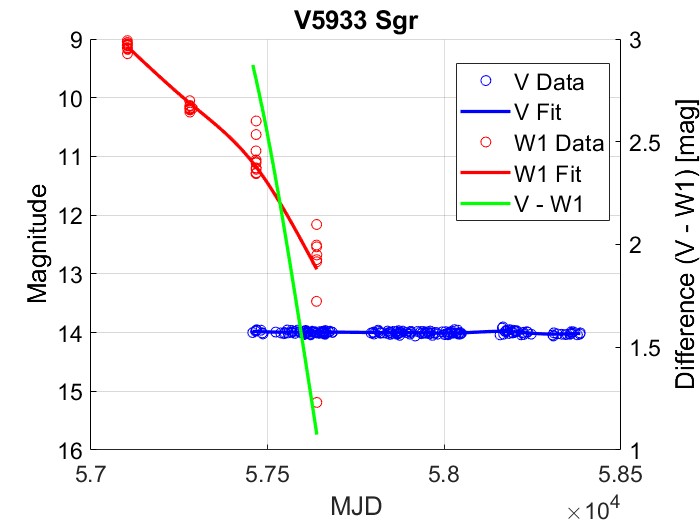} \\
\includegraphics[width=0.32\textwidth]{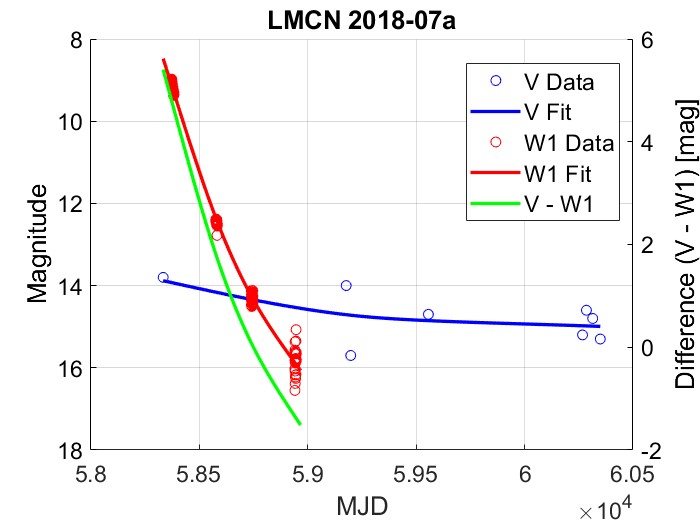} &

\includegraphics[width=0.32\textwidth]{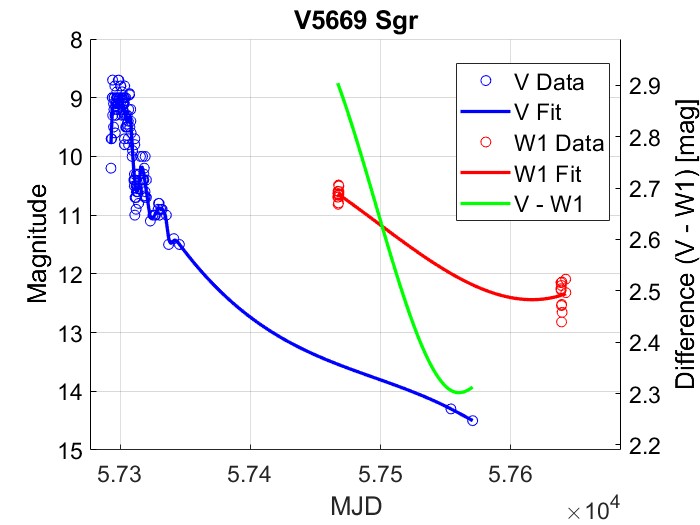} &
\includegraphics[width=0.32\textwidth]{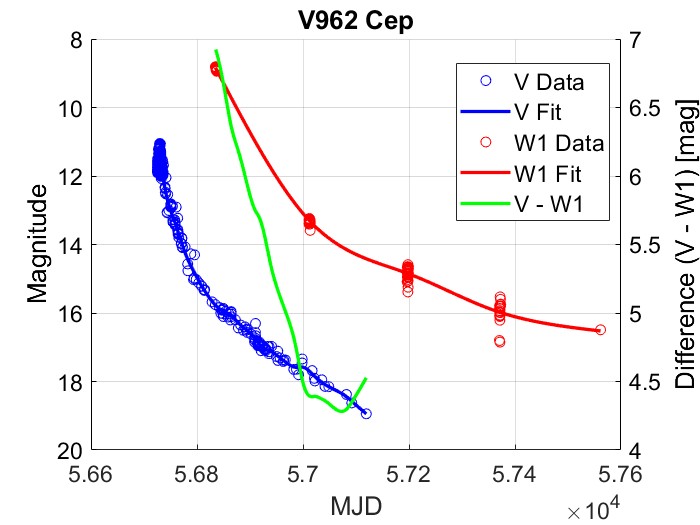} \\
\includegraphics[width=0.32\textwidth]{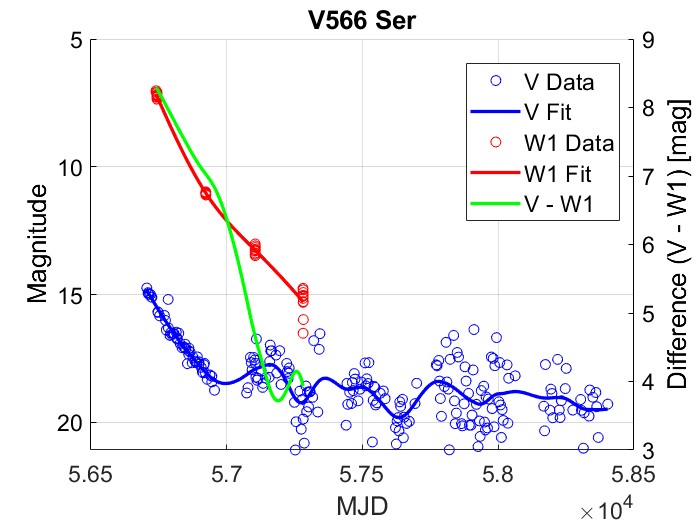} 
\end{tabular}
\caption {Continued:{ $V-W1$ color difference model applied to our identified samples of known novae. Individual data points are shown in red (WISE W1 band) and blue (optical V band), capturing the temporal evolution of each emission component. The green curve represents the computed ($V-W1$) color difference, which serves as a diagnostic of the nova’s emission characteristics.}}
    \label{fig:V-W1}
\end{figure*}

\begin{figure*}[t!]
\centering
\begin{tabular}{@{}ccc@{}}
\includegraphics[width=0.34\textwidth]{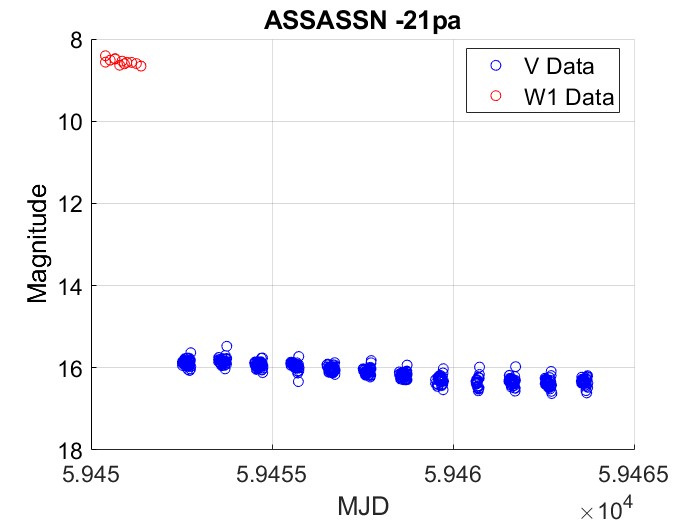} &
\includegraphics[width=0.34\textwidth]{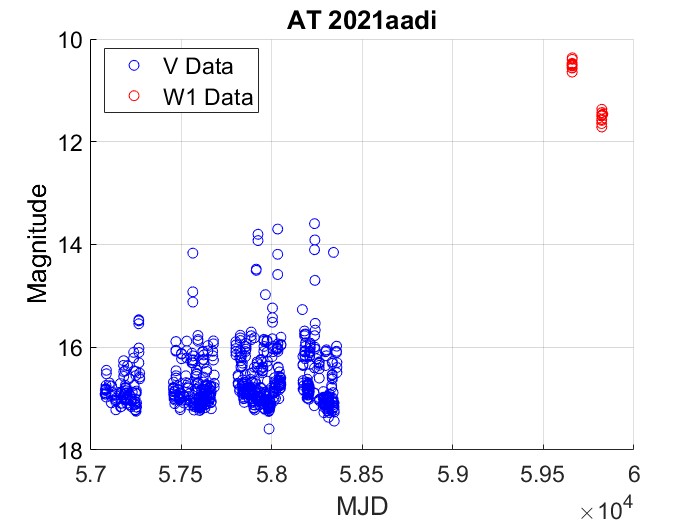} &
\includegraphics[width=0.34\textwidth]{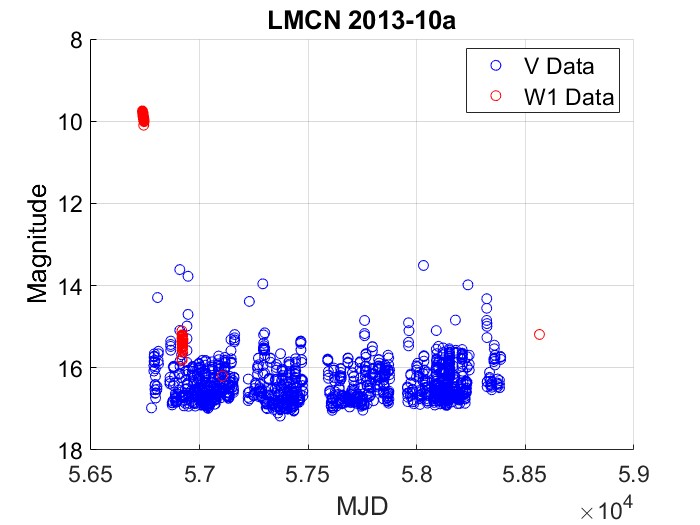} \\
\includegraphics[width=0.34\textwidth]{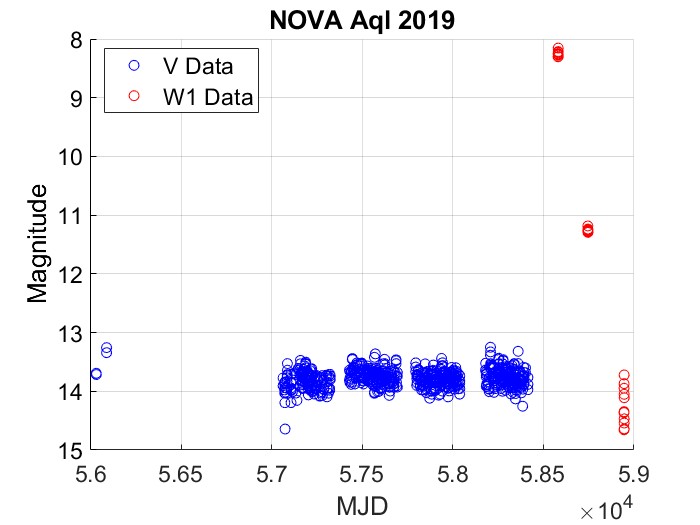} &        
\includegraphics[width=0.34\textwidth]{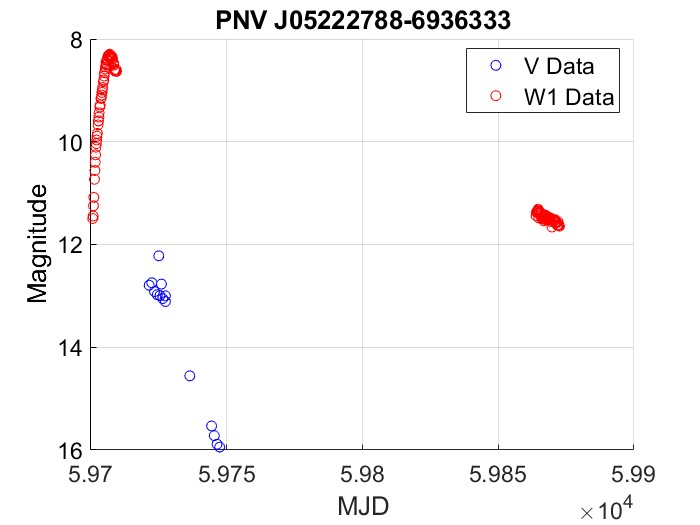} &
\includegraphics[width=0.34\textwidth]{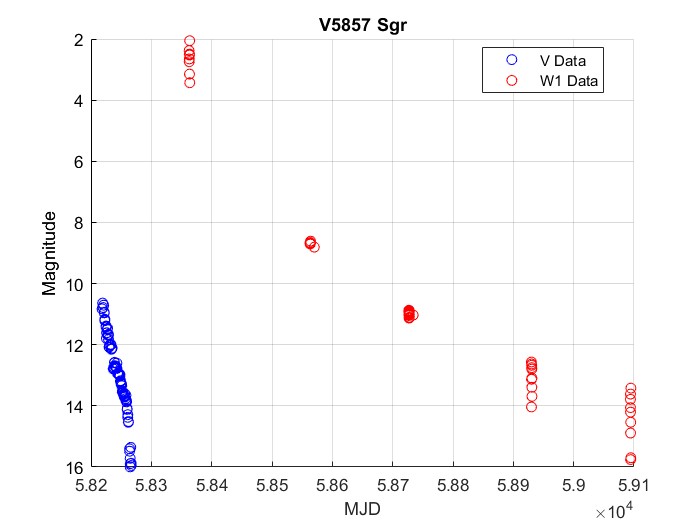} \\
\includegraphics[width=0.34\textwidth]{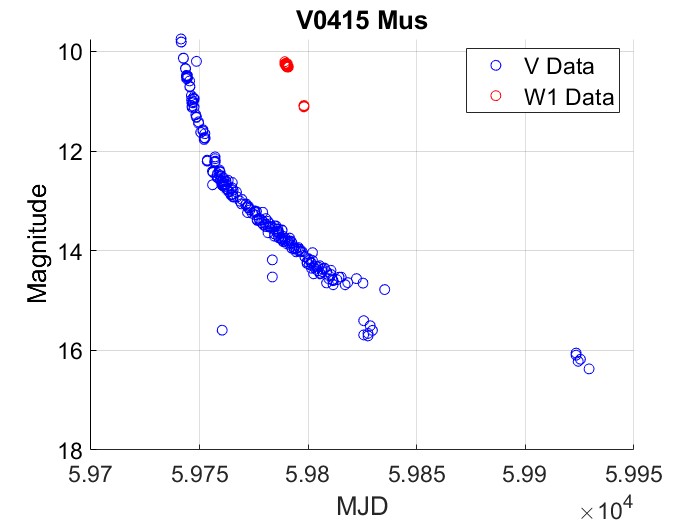} &
\includegraphics[width=0.34\textwidth]{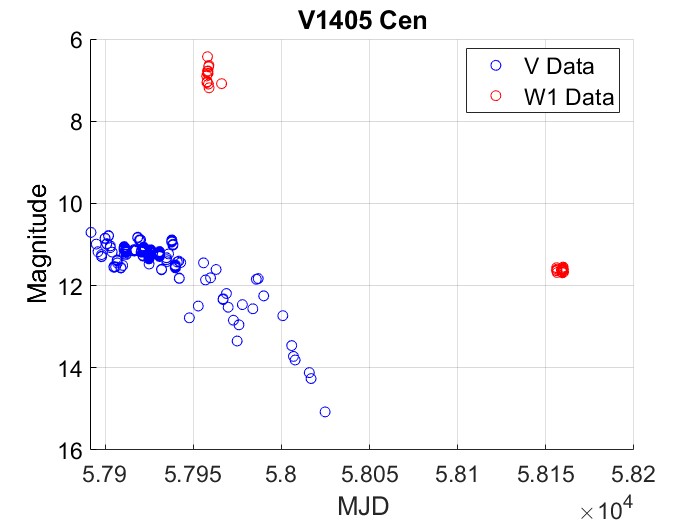} &
\includegraphics[width=0.34\textwidth]{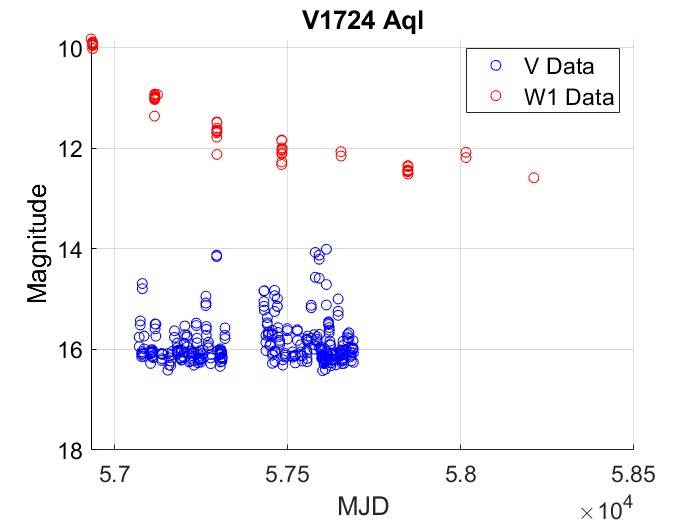} \\
\includegraphics[width=0.34\textwidth]{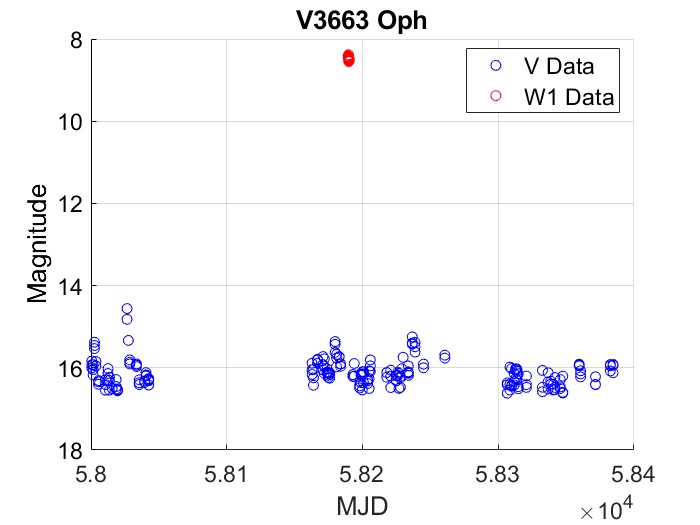} &
\includegraphics[width=0.34\textwidth]{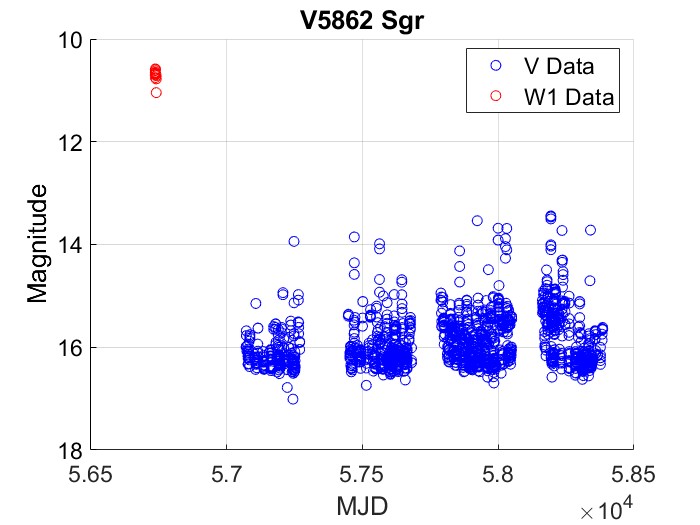} &
\includegraphics[width=0.34\textwidth]{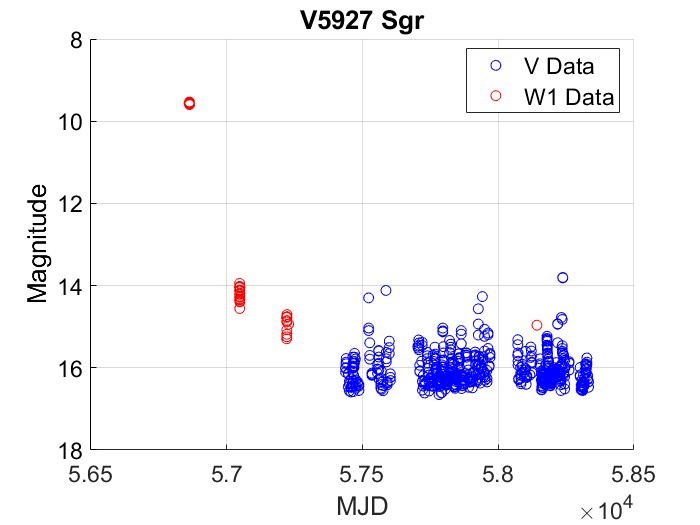} \\
\end{tabular}
\caption{Novae from our pipeline without suitable overlap between V and W1 for computing a $V-W1$ color difference. Individual data points are shown in red (WISE W1 band) and blue (optical V band), capturing the temporal evolution of each emission component.}
\label{fig:V-W11A}
\end{figure*}

\begin{figure*}[t!]
\centering
\begin{tabular}{ccc}
\includegraphics[width=0.33\textwidth]{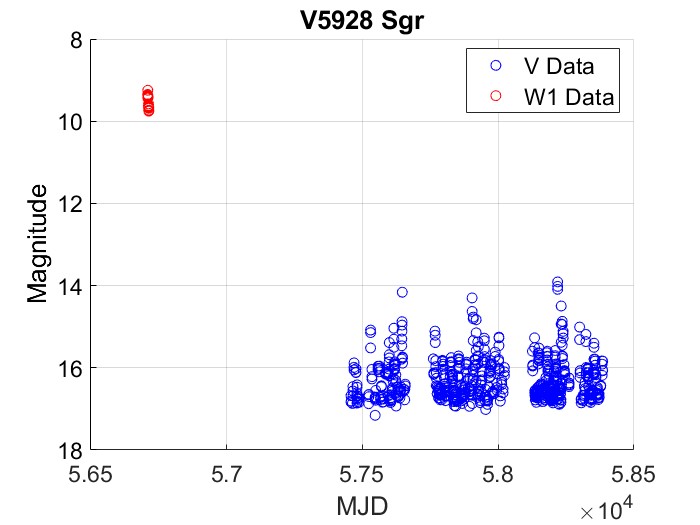} &
\includegraphics[width=0.33\textwidth]{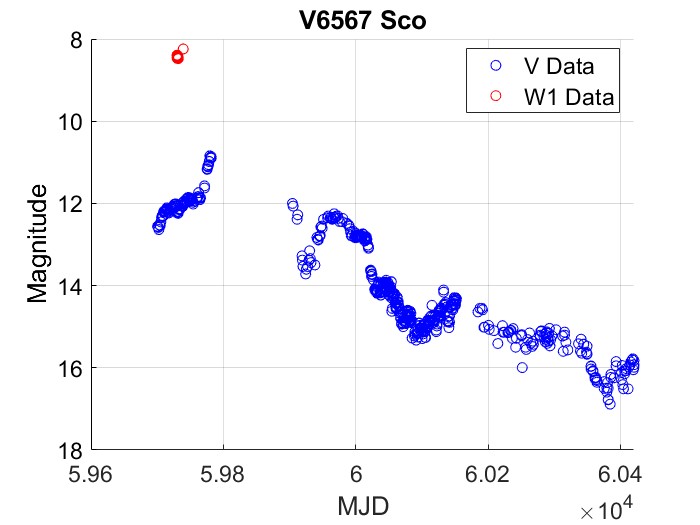} &
\includegraphics[width=0.33\textwidth]{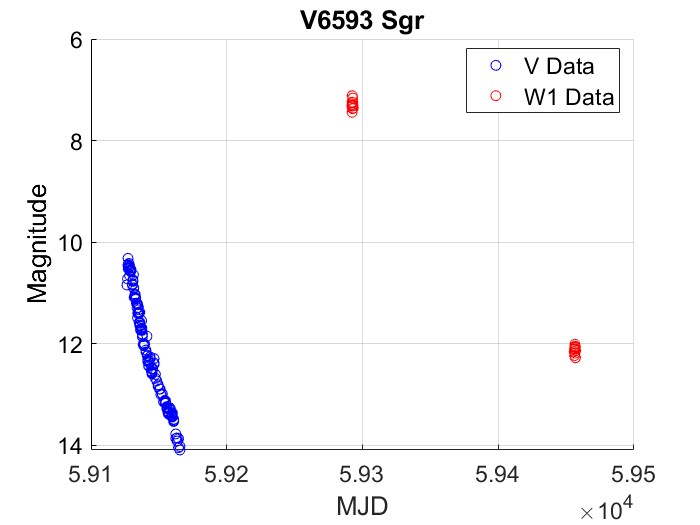} \\
\includegraphics[width=0.33\textwidth]{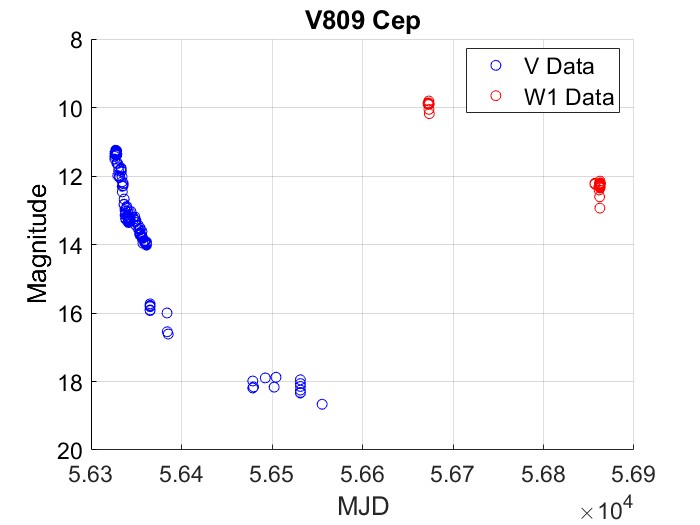} & & \\
\end{tabular}
\caption{Continued: Novae from our pipeline without suitable overlap between V and W1 for computing a $V-W1$ color difference. Individual data points are shown in red (WISE W1 band) and blue (optical V band), capturing the temporal evolution of each emission component.}
\label{fig:V-W11}
\end{figure*}
\FloatBarrier
 
\begin{figure*}[t!]
    \centering
{\includegraphics[width=1\textwidth]{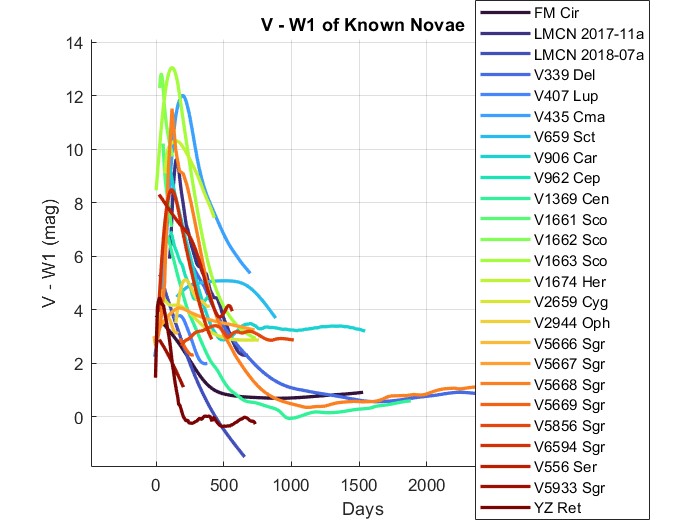}}
   \caption{$V-W1$ color evolution as a function of normalized time (in days) for the entire sample of nova candidates. Each line corresponds to an individual nova, with distinct colors assigned to differentiate between objects. The time axis has been normalized relative to each nova’s eruption time to facilitate direct comparison of their early color evolution.}
   \label{fig:MagVsTime}
  \end{figure*}
  
\begin{figure*}[t!]
    \centering
      {\includegraphics[width=0.8\textwidth]{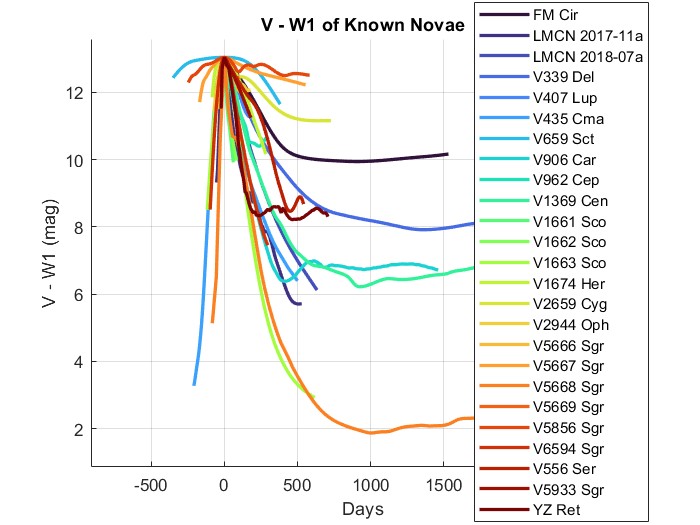}}
    \caption{$V-W1$ at fixed mag. vs Decline Time (days) of our novae samples. This shows the comprehensive color evolution of $V-W1$ for all 25 nova candidates included in our analysis, capturing the early and late evolution phases with a fixed peak to better understand their rate of decline. Each line represents a different nova candidate, with various colors denoting specific objects, allowing for easy differentiation and analysis of color evolution trends among our sample.}
    \label{fig:Declineplot}
\end{figure*}
\begin{figure*}[t!]
    \centering
\includegraphics[width=1\textwidth,height=1.2\textheight,keepaspectratio]{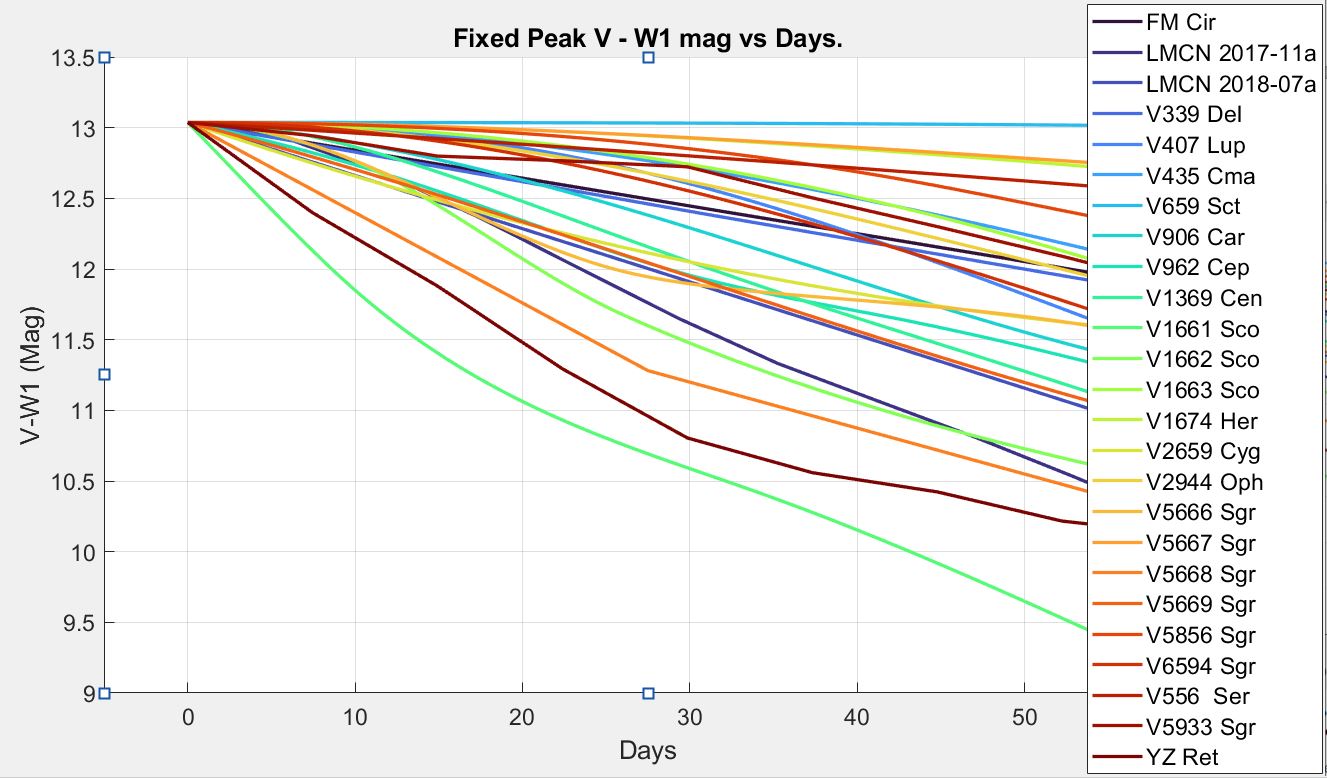}
\caption{$V-W1$ color evolution during the first 60 days post-maximum evolution for all nova candidates in our sample. The color index values are plotted against days after peak, with each line representing a distinct nova. Different colors are used to visually distinguish individual objects. By focusing on this early post-eruption phase, the plot captures the linear decline in $V-W1$ color before the onset of significant dust formation, emphasizing the intrinsic emission characteristics of the nova ejecta across the sample.}
    \label{fig:Declineplot2}
\end{figure*}

\begin{figure*}[t!]
    {\includegraphics[trim={0.0cm 0.0cm 0.0cm 0.9cm}, clip, width=2.00\columnwidth]{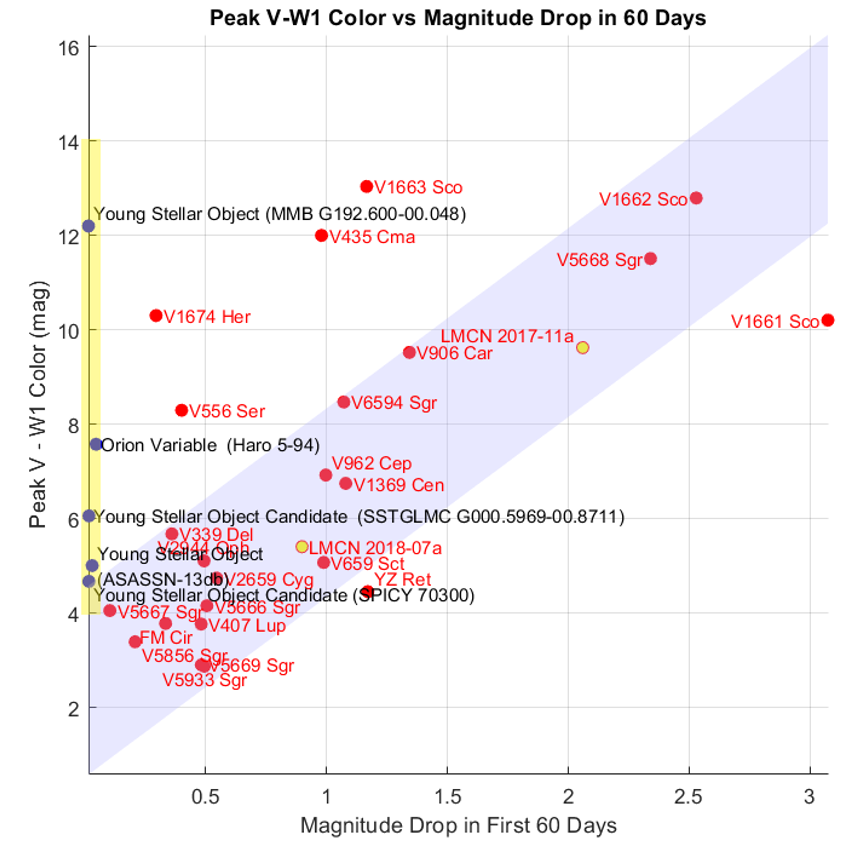}}
   \caption{Peak $V-W1$ vs. its magnitude decrease over the first 60 days. Red circles represent individual novae, labeled by name. The blue circles are the non-novae candidates from our samples.}
    \label{fig:DecVsPeak}
\end{figure*}

\begin{figure*}[t!]
    \centering
      {\includegraphics[trim={0.0cm 0.cm 0.0cm 0.9cm}, clip, width=1.99\columnwidth]{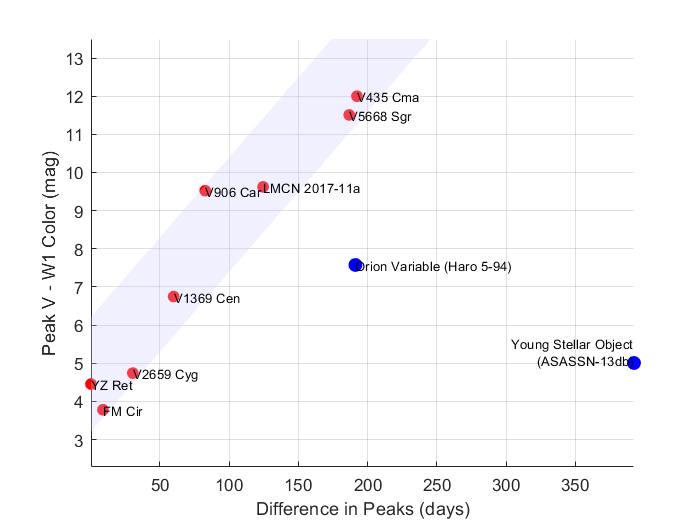}}
   \caption{ Peak $V-W1$ color vs. the time difference between the peaks of the $V$ and $W1$ light curves.} 
   \label{fig:PeakDiff}
\end{figure*}

\begin{figure*}[t!]
\centering
\resizebox{\textwidth}{0.30\textheight}{%
\begin{tabular}{ccc}
\includegraphics[width=0.45\textwidth]{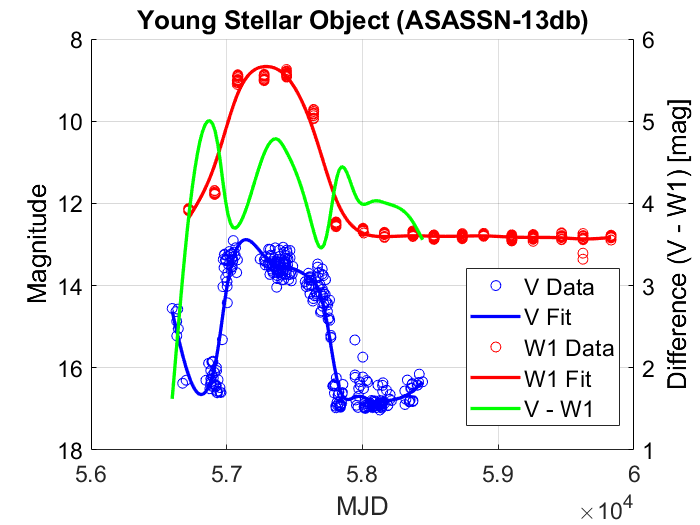} &
\includegraphics[width=0.45\textwidth]{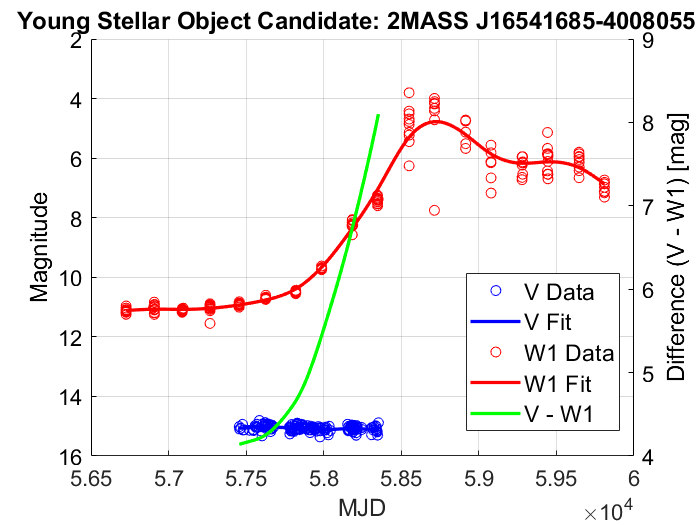} &
\includegraphics[width=0.45\textwidth]{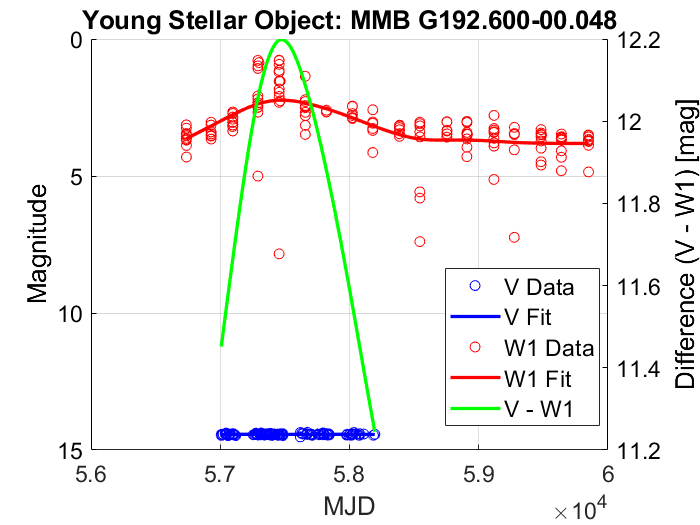} \\
\includegraphics[width=0.45\textwidth]{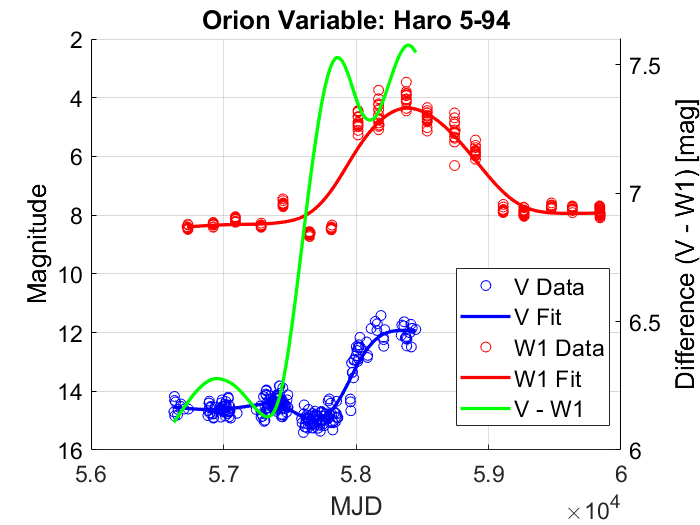} &
\includegraphics[width=0.45\textwidth]{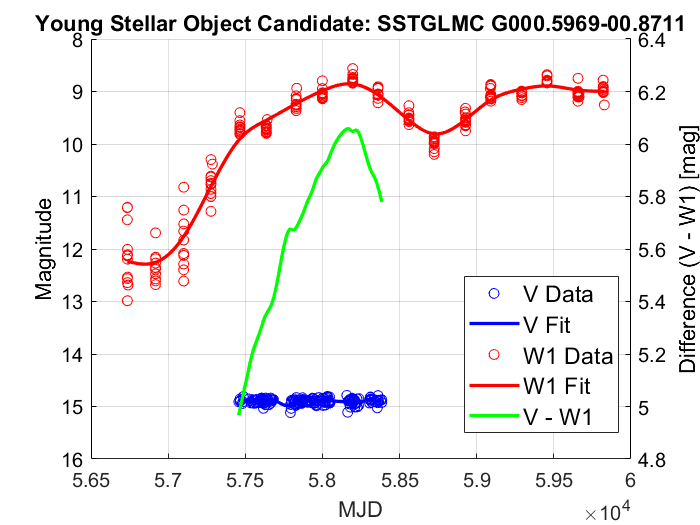} &
\includegraphics[width=0.45\textwidth]{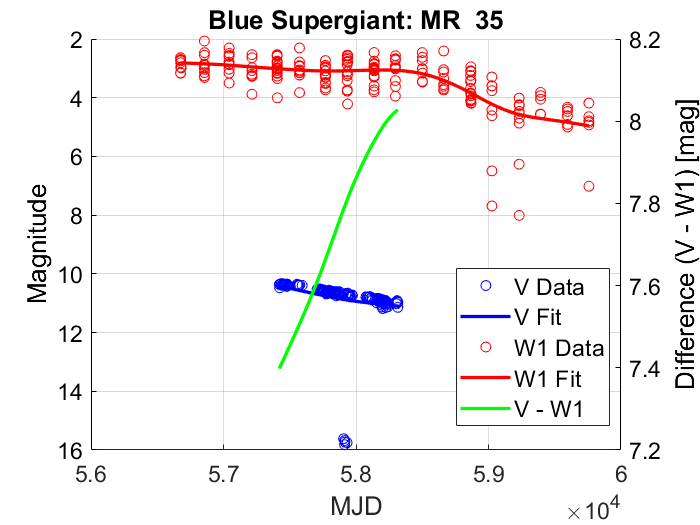} \\
\includegraphics[width=0.45\textwidth]{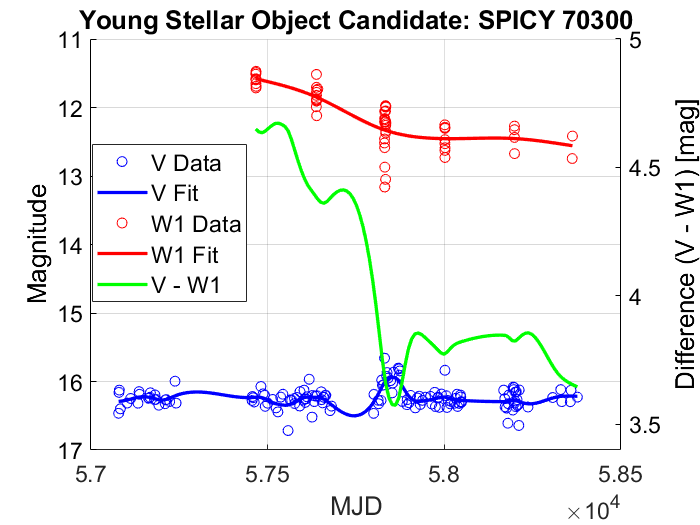} & & \\
\end{tabular}}
\caption{$V-W1$ color difference for optically known non-novae sources from our cross-match.} 
\label{fig:V-W3}
\end{figure*}

\subsection{V-W1 peak color vs. 60 day mag. change post-peak}
 
We constructed a plot of the peak $V-W1$ color index vs. the magnitude change 60 days after this color peak, for our 25 confirmed novae with overlapping optical and infrared observations. Our analysis reveals that novae with more rapid declines in the $V-W1$ color evolution systematically present larger, redder $V-W1$ colors at peak, whereas novae with slower declines display bluer colors as seen in Figure \ref{fig:DecVsPeak}. Quantitatively, our statistical analysis yields a highly significant, strong positive Pearson correlation coefficient of r = 0.69, indicating that as the $V-W1$ color evolves more rapidly, their $V-W1$ color is redder at peak and also that slower-evolving novae are bluer (smaller $V-W1$). 
This correlation suggests cooler ejecta in faster-declining novae and possibly different emission mechanisms, such as enhanced free-free emission processes. The redder colors in fast-declining novae could arise from enhanced free-free emission, while the bluer colors in slow novae may indicate prolonged contributions from recombination lines and bound-bound transitions in the optical \citep{1998PASP..110....3G,banerjee2013nearinfraredpropertiesclassicalnovae,Evans_2016}. 

The observed trend that faster novae exhibit redder, hence cooler, photospheres (see Fig. 8) can be interpreted in terms of ejecta energetics and expansion dynamics. A more energetic ejection drives a faster expansion, which in turn produces a larger photospheric radius and a more diluted radiative flux, resulting in a faster cooling rate. Consequently, the photosphere reaches lower temperatures earlier, shifting the spectral energy distribution (SED) peak toward longer wavelengths and making the nova appear redder at earlier epochs. This behavior can be quantified through the evolution of the ejecta optical depth:
\begin{align}
\tau(t) &= \kappa \frac{M_{\text{ej}}}{4 \pi R(t)^2} 
= \frac{\kappa M_{\text{ej}}}{4 \pi v^2 t^2}
\label{eq:1}
\end{align}

where $\kappa$ is the opacity and $R(t) \simeq vt$ describes the homologous expansion. The transition to the optically thin regime ($\tau = 1$) occurs at time $t_{\text{thin}}$, when the ejecta becomes optically thin:

\begin{align}
t_{\text{thin}} 
&= \sqrt{ \frac{\kappa M_{\text{ej}}}{4 \pi v^2} } 
\propto \sqrt{ \frac{M_{\text{ej}}}{v} }
\label{eq:2}
\end{align}

\bigskip 

showing that novae with higher expansion velocities (and lighter envelopes) become transparent earlier. Since the bolometric luminosity scales as:

\begin{align}
L = 4 \pi R^2 \sigma T^4,
\end{align}
and assuming that the luminosity during expansion is roughly constant (and equal to the Eddington luminosity) we find that the temperature evolves as:
\begin{align}
T(t) \propto v^{-1/2} t^{-1/2}
\end{align}

indicating that novae that decline faster, and novae with higher ejecta velocities exhibit lower photospheric temperatures, providing a natural explanation for their redder observed colors.
This interpretation is consistent with theoretical models in which low accretion rates lead to the accumulation of more massive envelopes before ignition \citep{1985ApJ...291..812K,1995ApJ...445..789P,2013ApJ...777..136W}, resulting in stronger cooling and redder colors. Conversely, systems with high accretion rates ignite smaller envelopes, producing less massive but hotter ejecta that appear bluer \citep{1982ApJ...257..752F,Hillman_2016,Starrfield_2016}. Overall, the scaling relations (Equations \ref{eq:1}--\ref{eq:2}) reinforce the idea that the early color evolution of novae to be governed primarily by expansion and radiative cooling of the ejecta. 

This interpretation is also consistent with models predicting that the early color evolution of novae is governed by ejecta expansion and cooling, especially in systems hosting high-mass white dwarfs \citep{1986ApJ...310..222P,2005ApJ...623..398Y,2011A&A...533L...8S,2016ApJ...819..168H,2018MNRAS.474.2679A,2022MNRAS.514.2239S}.
In fast novae, rapid cooling and changes in emission mechanism drive a spectral redistribution of the emergent flux, producing the systematically red $V-W1$ colors observed near peak. This reinforces the interpretation that the underlying physics operates in a self-similar way across different nova systems and that deviations in color are directly related to the dynamical evolution of the ejecta.
This observational signature supports the scenario that, in fast novae, the spectral energy distribution (SED) shifts toward the IR. 
\subsection{V-W1 peak vs. peak time difference}
Of the 25 confirmed novae for which we produced a $V-W1$ color curve, 8 have coverage for both the optical and W1 peak magnitude individually. For these 8 novae, we calculated the time difference between the two peaks (that of the $V$ band and that of the $W1$ band), and show in Figure \ref{fig:PeakDiff}, the peak $V-W1$ vs the time difference between the peaks in the two bands. This revealed a clear trend with all novae residing in a diagonal strip, indicating that novae that exhibit a cooler color curve peak have a longer time delay between the $V$ and $W1$ peaks. The optical peak of a nova marks the beginning of significant mass loss immediately following the TNR and the pre-maximum UV flash \citep{2011ApJS..197...31S,2011A&A...533L...8S,2014MNRAS.437.1962H}. Subsequently, the envelope rapidly expands, causing the optical decline and IR rise that produce the $V-W1$ color peak when the photospheric temperature reaches its minimum.

We can therefore say that the time difference between the optical peak and the IR peak represents the mass-loss timescale, which is the duration over which most of the envelope is ejected. \citet{2014MNRAS.437.1962H} show that the bolometric peak of a nova corresponds to the onset of mass loss, while the coolest point marks the maximum envelope expansion. The interval between these points encompasses the  mass ejection phase. Once mass ejection is complete, the envelope contracts and heats, shifting the emission back to shorter wavelengths. 

This physical interpretation explains our observed correlation in which novae with redder $V-W1$ peaks experienced more massive ejection events that required longer timescales to expel the material, resulting in extended delays between optical and IR peaks. This delay provides a direct photometric diagnostic of the mass-loss duration. Knowing the mass-loss timescale is crucial for understanding the total ejected mass, the efficiency of the TNR, and the likelihood of dust formation in the ejecta.
Notably, while all novae fall within a tight band in this plot, highlighting the consistency of this relationship, additional non-nova sources, plotted in blue, deviate significantly from this trend. This clear separation shows that the $V-W1$ color and IR delay are effective for identifying novae as well as distinguishing novae from other variables or young stellar objects.

In general, the observed strong positive correlation tells us that when the nova ejecta take longer to cool, the ejecta's thermal emission shifts to longer wavelengths, increasing the $V-W1$ color (redder). Previous studies also note that such delays in IR peak times reflect the time scales for the ejecta to cool, which depend on physical parameters like ejecta mass and expansion velocity \cite[]{ 2012BASI...40..213E,2017MNRAS.469.1314D}.

\begin{figure*}[t!]
    \centering
      {\includegraphics[width=1.0\textwidth]{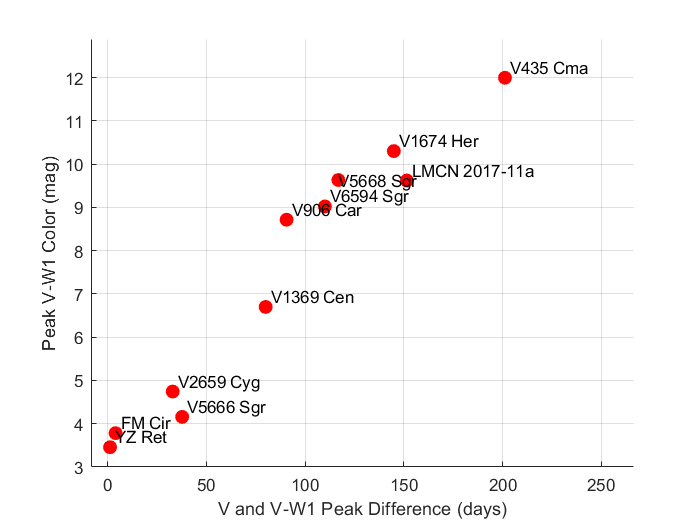}}
   \caption{Peak $V-W1$ color vs. the time lag between the peaks of V band and V-W1 color curve.}
 \label{fig:PeakVandVW1}
\end{figure*}

\subsection{V-W1 of identified non-novae from our catalog}
As shown in Figure \ref{fig:V-W3}, the $V-W1$ color evolution of the non-nova sources is markedly different from that of the confirmed novae presented in Figures \ref{fig:V-W1A} and \ref{fig:V-W1}. This is another important outcome of this study; it confirms the distinct photometric signatures of novae in the optical-mid-IR regime.

This observation prompts a critical question: are the correlations we observe in Figures \ref{fig:DecVsPeak} and  \ref{fig:PeakDiff} unique to novae, or could other transient classes exhibit similar photometric behavior?
To address this, we constructed a carefully selected comparison sample of non-nova sources recovered through our search methodology. We included seven such sources in Figure \ref{fig:DecVsPeak}, which plots peak $V-W1$ against the magnitude change within the first 60 days post-maximum, and three additional sources in Figure \ref{fig:PeakDiff}, comparing peak $V-W1$ with the time difference between optical and IR peaks. Selection criteria required sufficient simultaneous $V$ and $W1$ coverage to accurately compute $V-W1$ colors, and observable peaks in both spectral regimes for Figure \ref{fig:PeakDiff}.
Our analysis reveals that non-nova sources occupy distinctly separate regions in both parameter spaces compared to  novae. In Figure \ref{fig:DecVsPeak}, while slow novae cluster toward the upper right and fast novae toward the lower left, non-novae fall outside this well-defined distribution. Similarly, in Figure \ref{fig:PeakDiff}, their placement shows significant divergence from the nova population. 
This clear separation proves the diagnostic power of $V-W1$ color evolution: the characteristic photometric behavior of novae in the optical-mid-IR domain is not replicated by other transient populations. The inclusion of these non-nova comparison objects further strengthens our conclusion that $V-W1$ evolution serves as a strong tool for nova identification, providing a reliable, \textit{purely photometric} method for distinguishing novae from other transients.

\section{Discussion}
In the $V-W1$ color versus 60-day magnitude drop parameter space, novae are found to cluster within a relatively well-defined locus, reflecting their characteristic photometric evolution and color behavior.
The relationship illustrated in Figure \ref{fig:DecVsPeak} shows this correlation between the peak color of $V-W1$ and the decrease in magnitude, suggesting that a deeper understanding of the parameters of the nova system can improve our understanding of their behavior after eruption.
Further studies could explore the physical mechanisms driving this correlation, potentially leading to a more comprehensive model that accounts for various influencing factors in nova systems.

Our analysis of the time difference between the $V$ and $W1$ peaks and the peak $V-W1$ color, as shown in Figure \ref{fig:PeakDiff} confirms a fundamental link between the nova's evolutionary timescale and the physical properties of its ejecta. Fast novae with short infrared delays show bluer $V-W1$ colors, indicative of less massive eruptions where the ejecta maintains a higher effective temperature due to a smaller expansion radius, whereas slower novae or those with extended ejecta structures display longer delays and redder colors. This corresponds to more energetic eruptions with larger ejected masses.
This correlation is driven by the intrinsic properties of the expanding photosphere, where the $V-W1$ peak color and the time delay between the $V$ and $W1$ peaks serve as direct indicators for the mass and energy of the eruption.

All novae fall within a tight band in this correlation, whereas additional non-nova sources plotted in the same parameter space deviate significantly from the trend. This  distinction indicates that the $V-W1$ color evolution and IR delay provide a robust diagnostic for identifying novae and distinguishing them from other transients.

An additional significant result from our analysis reveals that the decline rate of the $V-W1$ color curve for young stellar objects (YSOs) 
is very slow, and does not depend on the peak $V-W1$ color evolution --- in stark contrast to novae behavior.

The absence of such correlation within the YSO population suggests fundamentally different physical processes driving the observed variability. For novae, the $V-W1$ evolution is intrinsically linked to the TNR event, producing a distinct photometric signature that scales with the eruption properties. In contrast, the relatively uniform $V-W1$ decline behavior observed in YSOs likely originates from mechanisms such as accretion variability, circumstellar disk obscuration events, or changes in dust emission properties, none of which would be expected to correlate systematically with the initial optical to IR color difference.

In Figure \ref{fig:PeakVandVW1}, we plot the peak $V-W1$ color against the time delay between the optical $V$ band peak and the $V-W1$ color peak for candidates for which we have data for both of these peaks. The figure reveals a strong positive correlation, where novae with a longer time delay also exhibit a significantly redder (cooler) peak $V-W1$ color. For example, fast evolving novae like YZ Ret and FM Cir show short delays of less than 10 days and peak colors below 4 mag, while slower-evolving systems like V435 Cma and V1674 Her have delays exceeding 140 days and peak color differences of greater than 10 mag.
This relationship provides a powerful diagnostic for the physical timescale of nova eruptions. The time delay can be interpreted as the duration of the primary mass loss and envelope expansion phase. Novae with more energetic eruptions eject larger amounts of mass, causing the photosphere to expand to a greater radius. This results in both a cooler, redder peak color and a more extended period of mass ejection, leading to the longer observed time delay. This strongly indicates that the peak $V-W1$ color is a direct tracer of the eruption's energy and the mass of the ejected envelope, with the time delay between the optical peak and the $V-W1$ color evolution peak serving as a measure of the event's evolutionary timescale.

\section{Conclusions} 

Using the light curve characteristics of the confirmed novae recovered by our pipeline as a reference, we performed a visual analysis of the light curves of the remaining about 1,800 unmatched sources. This process yielded a catalog of 58 high-confidence objects exhibiting strong nova behavior. Combining these results, we obtained a catalog of 113 sources of interest: 41 optically confirmed novae and 72 candidates.  These findings may provide an independent way to estimate the Galactic nova rate, which remains uncertain, with estimates ranging from about 25 novae $\rm yr^{-1}$ up to nearly 150 novae $\rm yr^{-1}$. Because most historical nova surveys in the Milky Way have been conducted in the optical domain, they are likely affected by severe incompleteness due to interstellar extinction, particularly toward the Galactic plane. IR observations can mitigate this bias, offering a less extinction-sensitive view of the nova population. If we assume that all unclassified variable sources identified by our pipeline are indeed novae, the correction factor to apply to optically derived nova rates could be as high as about $41+72/41\sim 2.8$. However, we note that not all 72 unclassified variables are necessarily novae; hence, this value should be regarded as a robust upper limit. A more realistic correction can be expressed as $(41+g\times 72)/41$ where $g$ is the fraction of true novae among the candidates. Previous analyses of time-domain surveys suggest that the "purity" of nova candidate samples is typically modest: for instance, \citet{2015ApJS..219...26M,2016ApJS..222....9M} find that only about 20-40\% of initially selected “candidate” variables are confirmed as genuine novae after follow-up. Therefore, the correction factor would lie in the range 1.4--1.8. Classical nova rates based on optical surveys alone, ranging from $\sim$20 to 30 novae yr$^{-1}$ \cite[e.g., ][]{1957gano.book.....G, 1994A&A...286..786D}, would thus increase to $\sim$40--50 novae yr$^{-1}$ after accounting for incompleteness, in line with modern estimates or derived from combined optical and IR surveys \cite[][for a review and discussion]{2017ApJ...834..196S,2022ApJ...936..117R,2020A&ARv..28....3D}.
Our analysis, therefore, provides an independent estimate of the Galactic nova rate, highlighting the importance of IR monitoring in revealing systems that remain undetected in the optical due to interstellar extinction. The inferred rate of $\sim$40--50 novae yr$^{-1}$ supports the view that infrared time-domain surveys can significantly mitigate the long-standing incompleteness affecting optical searches, offering a more comprehensive census of nova eruptions across the Galaxy.

The relationship we observe between peak $V-W1$ and the decline rate of the early light curve confirms theoretical predictions that more massive ejection events produce cooler photospheres due to greater envelope expansion \citep{1986ApJ...310..222P,2009ApJ...704.1676K,2005ApJ...623..398Y,2013ApJ...777..136W}. Models predict that systems with lower accretion rates accumulate larger envelope masses before ignition, leading to more energetic explosions that drive the envelope to larger radii, thereby decreasing the effective temperature \citep{2005ApJ...623..398Y,Starrfield_2016}. Our findings indicate that a higher $V-W1$ peak corresponds to a more expanded envelope, signifying a stronger eruption that ejects more mass. Since these eruptions are more energetic, they expel material at higher velocities, causing the mass-loss episode to conclude more quickly and the envelope to shrink faster, thereby explaining the rapid color decline observed for novae with cooler (redder) peaks. The correlation with peak delay reflects the combined timescales of envelope expansion and mass ejection, which are longer for more massive ejection events \citep{1995ApJ...445..789P,2016ApJ...819..168H,Shara_2018}.

This work emphasizes the importance of conducting coordinated multi-band observations in optical and mid-IR bands from the earliest phases of nova eruptions. Such observations enable rapid classification that can trigger appropriate follow-up strategies: fast novae with red peaks and steep declines should be prioritized for high-energy observations, as their more energetic explosions and faster ejecta are more likely to produce shocks capable of particle acceleration \citep{2015MNRAS.450.2739M,2014Natur.514..339C,2018A&A...609A.120F,2020NatAs...4..776A}. Indeed, all gamma-ray detected novae to date have been relatively fast novae with $t_2 <$ 12 days \citep{2014Sci...345..554A,2016ApJ...826..142C}. Conversely, slow novae with blue peaks may be better candidates for long-term optical monitoring to study their complex light curve evolution and potential secondary peaks \citep{2010AJ....140...34S}.

Regarding dust formation, our analysis suggests that dust primarily affects the light curve after the $V-W1$ peak, consistent with the requirement that the ejecta must first be fully expelled and then cool below condensation temperatures (~1500 K). This explains why the early color evolution (first 60 days) primarily reflects the intrinsic properties of the expanding photosphere rather than dust effects.

This work introduces the $V-W1$ color evolution as a new diagnostic tool for the early stages of nova eruptions. By measuring the peak $V-W1$ color and its decline rate within the first couple of months, we can not only classify nova types, distinguishing fast novae (steep color decline, red peak) from slow novae (gradual color decline, blue peak), but also identify newly detected transients as novae and differentiate them from other types of transient events. The color-time mappings shown in Figures 8, 9, and 11 provide a practical framework for applying this method to ongoing and future transient surveys, such as  LSST, where the volume of discoveries is expected to exceed spectroscopic follow-up capacity \citep{2010AJ....140...34S,2017Natur.548..558S,2022MNRAS.514.2239S,Kisley_2023}.
 
\section*{Acknowledgments} 
Ariel University School of Graduate Studies Research Scholarship funded this research work. This work was supported by the Azrieli College of Engineering -- Jerusalem Research Fund.
We acknowledge the use of data from SIMBAD and WISE databases.  We acknowledge with thanks the variable star observations from the AAVSO International Database contributed
by observers worldwide and used in this research. 

\bibliography{sample631}{}
\bibliographystyle{aasjournal}

\end{document}